\documentclass[final,5p,times,sort,compress]{elsarticle} 

\usepackage{lineno}
\usepackage{graphicx} 
\usepackage{amsmath}
\usepackage{caption}
\usepackage{subcaption}
\usepackage{tikz,pgfplots}
\usetikzlibrary {patterns,patterns.meta}
\usepackage{hyperref}
\usepackage{todonotes}
\usepackage{amssymb}
\usepackage{multirow}
\pgfplotsset{compat=1.18}
\usepackage{bm}
\usepackage{cleveref}

\DeclareMathOperator*{\argmin}{\mathrm{argmin}}

\newcommand{\RR}{\mathbf{R}}

\newcommand{\AT}{\textbf{A}^\mathrm{T}}

\newcommand{\taulower}{\tau_\mathrm{lower}}
\newcommand{\tauupper}{\tau_\mathrm{upper}}

\newcommand{\Omegaout}{\Omega_\mathrm{out}}
\newcommand{\Omegain}{\Omega_\mathrm{in}}

\newcommand{\IPDR}{\mathrm{IPDR}}
\newcommand{\PW}{\mathrm{PW}}
\newcommand{\VER}{\mathrm{VER}}

\newcommand{\OSPW}[1]{\mathrm{OSPW}(w#1)}

\newcommand{\etal}{~\textit{et~al.}}

\journal{Additive Manufacturing}

\begin{document}
\begin{frontmatter}
\title{Systematic Analysis of Penalty-Optimised Illumination Design for Tomographic Volumetric Additive Manufacturing via the Extendable Framework TVAM AID Using the Core Imaging Library}

\author[inst1, corref]{Nicole Pellizzon} \ead{nicpel@dtu.dk}
\author[inst2]{Richard Huber} 
\author[inst1]{Jon Spangenberg}
\author[inst3]{Jakob Sauer Jørgensen}

\affiliation[inst1]{organization={Department of Civil and Mechanical Engineering},
            addressline={Technical University of Denmark}, 
            city={Kongens Lyngby},
            postcode={2800}, 
            country={Denmark}}

\affiliation[inst2]{organization={IDea Lab},
            addressline={University of Graz}, 
            city={Graz},
            postcode={8010}, 
            country={Austria}}

\affiliation[inst3]{organization={Department of Applied Mathematics and Computer Science},
            addressline={Technical University of Denmark}, 
            city={Kongens Lyngby},
            postcode={2800}, 
            country={Denmark}}

\begin{abstract}
Tomographic Volumetric Additive Manufacturing~(TVAM) is a novel manufacturing method that allows for the fast creation of objects of complex geometry in layerless fashion. The process is based on the solidification of photopolymer that occurs when a sufficient threshold dose of light-energy is absorbed. In order to create complex shapes, an illumination plan must be designed to force solidification in some desired areas while leaving other regions liquid. Determining an illumination plan can be considered as an optimisation problem where a variety of objective functionals (penalties) can be used. This work considers a selection of penalty functions and their impact on selected printing metrics; linking the shape of penalty functions to ranges of light-energy dose levels in in-part regions that should be printed and out-of-part regions that should remain liquid. Further, the threshold parameters that are typically used to demarcate minimum light-energy for in-part regions and maximum light-energy for out-of-part regions are investigated systematically as design parameters on both existing and new methods. This enables the characterisation of their effects on some selected printing metrics as well as informed selection for default values. This work is underpinned by a reproducible and extensible framework, TVAM Adaptive Illumination Design~(TVAM AID), which makes use of the open-source Core Imaging Library~(CIL) that is designed for tomographic imaging with an emphasis on reconstruction. The foundation of TVAM AID which is presented here can hence be easily enhanced by existing functionality in CIL thus lowering the barrier to entry and encouraging use of strategies that already exist for reconstruction optimisation.

\end{abstract}




\end{frontmatter}

\section{Introduction}
\label{section:introduction}

The manufacturing technique Tomographic Volumetric Additive Manufacturing~(TVAM)~\cite{CAL,loterie,bernal2019volumetric} is based on the physical process of solidification of liquid photopolymer when exposed to light~\cite{shusteff2017one}. Most additive manufacturing methods~\cite{frazier2014metal,martin20173d,giannatsis2009additive,sandstrom2016non,murphy20143d}, construct the object one layer at a time, which can both negatively impact the manufacturing quality~\cite{bikas2016additive,monzon2017anisotropy} as well as lead to a long print time, creating a major bottleneck~\cite{moroni2018biofabrication}. In contrast, TVAM can produce the whole object simultaneously yielding immense potential for faster manufacturing~\cite{bernal2019volumetric}. 

\begin{figure}
     \centering
     \begin{tikzpicture}[scale=2]

\draw[fill,black] (0,0) circle (1 cm) node[xshift=-1cm,yshift=0.8cm,white]{$\Omegaout$};

\draw[] (0.8,0.8) node[] {$\Omega$};

\foreach \posix/\posiy/\axisx/\axisy/\myrotation in {0/0/0.5/0.3/0, 0.5/0.4/0.3/0.2/45,-0.3/-0.5/0.2/0.4/15, 0.5/-0.5/0.3/0.3/0}
{\draw[white,fill,rotate = \myrotation] (\posix,\posiy) ellipse (\axisx cm and  \axisy cm) node[ black]{$\Omegain$};
}

\draw[dashed, ->] (-1,-1) -- (1,-1) node[right]{$\mathbf{x}$};
\draw[dashed, ->] (-1,-1) -- (-1,1) node[above]{$\mathbf{y}$};

\draw[thick,white] (0,-1) -- (0,-1.27);

\end{tikzpicture}
     \caption{Decomposition of slice~$\Omega$, a continuous domain, into the solidified in-part~$\Omegain$ (white) and liquid out-of-part~$\Omegaout$ (black) regions. Note the circular shape of~$\Omega$ mimicking a cylindrical cuvette.}
     \label{Fig_illustration_in_and_out}
 \end{figure}
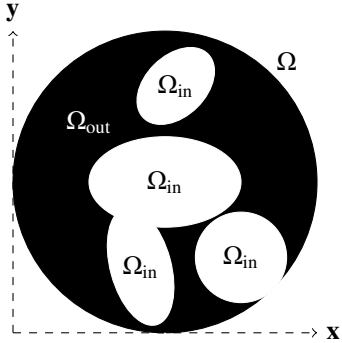

The gelation, or solidification, of the photopolymer occurs when the light-energy that is absorbed surpasses the threshold dosage~$\tau$. In TVAM, the degree of cure with respect to received light-energy dose is non-linear~\cite{CAL, highFidelity, highlyTunable} and could generally be considered to have two states~\cite{CAL, OSMO, highFidelity, loterie}: liquid when the accumulated dose is less than~$\tau$, and solidified when the accumulated dose is greater than~$\tau$. Thus, the printing domain~$\Omega$ is decomposed into the in-part region~$\Omegain \subset \Omega$, where solidification is desired, and the out-of-part region~$\Omegaout \subset \Omega$, where the photopolymer should remain liquid; see Figure~\ref{Fig_illustration_in_and_out} for an exemplary decomposition. For the common parallel projection systems~\cite{CAL, newGermanVAMGroup, loterie}, the illumination plans for different slices in the vertical height parameter~$\textbf{z}$ are completely independent and can be determined individually for each fixed~$\textbf{z}$. Using this model, the approaches introduced in Section~\ref{section_methodology} are described for a single slice~$\Omega \subset \RR^2$. However, the aim of TVAM is to create a 3D object where the whole volume is manufactured simultaneously, hence the `volumetric' designation in the name, which motivates the use of 3D voxels rather than 2D pixels. A discretised slice can then be considered to have dimension~$(m, n, 1)$. Application in 3D will be shown in~Section~\ref{section_3D}.

Achieving a suitable energy dose profile (analogous to a reconstructed image) to selectively cure specific voxels, it is necessary to plan how the photopolymer should be illuminated. However, as the illumination process affects all material along the rays of light that are cast into the photopolymer, it is impossible to illuminate a region without also illuminating the regions behind or in front. Indeed, if light were to be shone from just one single direction, neglecting refraction and assuming the light can penetrate the full distance, it would only be possible to create strip-shaped objects in the illumination direction. Such a paradigm does not offer sufficient flexibility to construct the complex geometries that are desired in practice.

To overcome this limitation of achievable printable geometry, the photopolymer is rotated so that illumination from all directions is possible (in practice the cuvette is often rotated multiple times to deliver sufficient light-energy for solidification to occur~\cite{highFidelity, VirtualVam, 3projectors}). The energy dose deposited by any single light ray is not sufficient to trigger the polymerisation reaction, rather, an accumulation from multiple sources at different angles is required. This allows voxels internal to the printing volume to be targeted without surpassing the energy dose in other voxels. Mathematically, the energy dose profile achieved by an illumination plan can be described by the back-projection operator (or the exponential back-projection to incorporate attenuation effects, also achievable through ray-tracing methods~\cite{rayDistortion, webber2023versatile}), which is strongly connected to the Computed Tomography~(CT) problem~\cite{natterer2001mathematics}.

Filtered Back-Projection (FBP) is a well-established reconstruction technique to find the ideal sinogram that, when back-projected, produces the desired image~\cite{natterer2001mathematics,deans2007radon}. However, there is no guarantee that a sinogram obtained via FBP will be non-negative; in most cases, the sinogram will attain negative values. Thus, such a sinogram is not a suitable illumination plan as the light source used in TVAM can only produce an emission of value equal to or greater than zero and hence, there is a natural non-negativity constraint on feasible illumination plans. Initial solutions proposed to project the negative parts of the sinogram to zero to create an illumination plan appropriate for TVAM~\cite{loterie, bernal2019volumetric}. But, this corrupts the image that is reconstructed (i.e. the energy dose profile) which in TVAM means the printed geometry no longer matches the desired geometry. This non-negativity constraint has, therefore, often been coupled with additional adjustments to the illumination plan in the form of iterative solvers~\cite{CAL, highlyTunable, OSMO, highFidelity, li2024tomographic, 3projectors} to improve the outcome. The gradient-based optimisation nature of these works is maintained here with specific inspiration drawn from two works in the literature: Rackson\etal~\cite{OSMO} with their use of upper and lower thresholds replacing the typical binary target geometry in optimisation algorithm Object Space Model Optimisation~(OSMO); and Wechsler\etal~\cite{waveOptical, inverseRendering} who formulated the dual threshold concept as a penalty function with additional penalty prescribed to over-polymerisation in their framework DrTVAM.

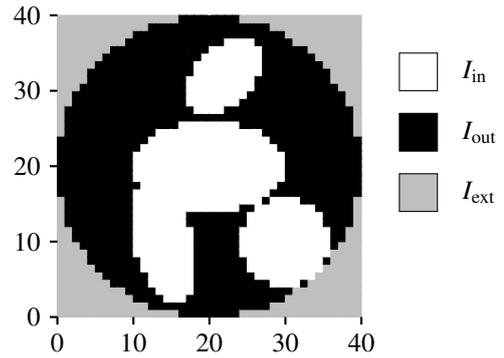
\begin{figure}
    \centering










\begin{tikzpicture}[scale=2]
  \def\xmin{-1} \def\xmax{1}
  \def\ymin{-1} \def\ymax{1}

  \def\npix{40} 
  \pgfmathsetmacro{\dx}{(\xmax-\xmin)/\npix}
  \pgfmathsetmacro{\dy}{(\ymax-\ymin)/\npix}

  \def\xa{0}   \def\ya{0}   \def\aa{0.5} \def\ba{0.3} \def\phia{0}
  \def\xb{0.1} \def\yb{0.6} \def\ab{0.3} \def\bb{0.2} \def\phib{45}
  \def\xc{-0.3}\def\yc{-0.5}\def\ac{0.2}\def\bc{0.4}\def\phic{15}
  \def\xd{0.5} \def\yd{-0.5}\def\ad{0.3} \def\bd{0.3} \def\phid{0}

  \foreach \i in {0,...,39}{
    \foreach \j in {0,...,39}{
      \pgfmathsetmacro{\cx}{\xmin + (\i + 0.5)*\dx}
      \pgfmathsetmacro{\cy}{\ymin + (\j + 0.5)*\dy}

      \def\pixcolor{gray!50}

      \pgfmathparse{(\cx*\cx + \cy*\cy) <= 1 ? 1 : 0}
      \ifnum\pgfmathresult=1
        \def\pixcolor{black}
      \fi

      \pgfmathsetmacro{\dxA}{\cx - \xa}
      \pgfmathsetmacro{\dyA}{\cy - \ya}
      \pgfmathsetmacro{\cA}{cos(\phia)}
      \pgfmathsetmacro{\sA}{sin(\phia)}
      \pgfmathsetmacro{\xAp}{\dxA*\cA + \dyA*\sA}
      \pgfmathsetmacro{\yAp}{-\dxA*\sA + \dyA*\cA}
      \pgfmathparse{(\xAp*\xAp)/(\aa*\aa) + (\yAp*\yAp)/(\ba*\ba) <= 1 ? 1 : 0}
      \ifnum\pgfmathresult=1
        \def\pixcolor{white}
      \fi

      \pgfmathsetmacro{\dxB}{\cx - \xb}
      \pgfmathsetmacro{\dyB}{\cy - \yb}
      \pgfmathsetmacro{\cB}{cos(\phib)}
      \pgfmathsetmacro{\sB}{sin(\phib)}
      \pgfmathsetmacro{\xBp}{\dxB*\cB + \dyB*\sB}
      \pgfmathsetmacro{\yBp}{-\dxB*\sB + \dyB*\cB}
      \pgfmathparse{(\xBp*\xBp)/(\ab*\ab) + (\yBp*\yBp)/(\bb*\bb) <= 1 ? 1 : 0}
      \ifnum\pgfmathresult=1
        \def\pixcolor{white}
      \fi

      \pgfmathsetmacro{\dxC}{\cx - \xc}
      \pgfmathsetmacro{\dyC}{\cy - \yc}
      \pgfmathsetmacro{\cC}{cos(\phic)}
      \pgfmathsetmacro{\sC}{sin(\phic)}
      \pgfmathsetmacro{\xCp}{\dxC*\cC + \dyC*\sC}
      \pgfmathsetmacro{\yCp}{-\dxC*\sC + \dyC*\cC}
      \pgfmathparse{(\xCp*\xCp)/(\ac*\ac) + (\yCp*\yCp)/(\bc*\bc) <= 1 ? 1 : 0}
      \ifnum\pgfmathresult=1
        \def\pixcolor{white}
      \fi

      \pgfmathsetmacro{\dxD}{\cx - \xd}
      \pgfmathsetmacro{\dyD}{\cy - \yd}
      \pgfmathsetmacro{\cD}{cos(\phid)}
      \pgfmathsetmacro{\sD}{sin(\phid)}
      \pgfmathsetmacro{\xDp}{\dxD*\cD + \dyD*\sD}
      \pgfmathsetmacro{\yDp}{-\dxD*\sD + \dyD*\cD}
      \pgfmathparse{(\xDp*\xDp)/(\ad*\ad) + (\yDp*\yDp)/(\bd*\bd) <= 1 ? 1 : 0}
      \ifnum\pgfmathresult=1
        \def\pixcolor{white}
      \fi

      \draw[fill=\pixcolor,draw=\pixcolor] 
        (\cx-0.5*\dx,\cy-0.5*\dy) rectangle (\cx+0.5*\dx,\cy+0.5*\dy);
    }
  }


\begin{scope}[shift={(\xmax+0.25,0)}]


  \draw[fill=white,draw=black] (0,0.5) rectangle +(0.25,0.25);
  \node[anchor=west] at (0.35,0.625) {$I_{\mathrm{in}}$};

  \draw[fill=black,draw=black] (0,0.1) rectangle +(0.25,0.25);
  \node[anchor=west] at (0.35,0.225) {$I_{\mathrm{out}}$};

  \draw[fill=gray!50,draw=black] (0,-0.3) rectangle +(0.25,0.25);
  \node[anchor=west] at (0.35,-0.175) {$I_{\mathrm{ext}}$};

\end{scope}

\foreach \k in {0,10,20,30,40}{
  \pgfmathsetmacro{\xtick}{\xmin + \k*(\xmax-\xmin)/\npix}
  \draw[thick] (\xtick,\ymin) -- (\xtick,\ymin-0.05);
  \node[below] at (\xtick,\ymin-0.05) {\k};
}

\foreach \k in {0,10,20,30,40}{
  \pgfmathsetmacro{\ytick}{\ymin + \k*(\ymax-\ymin)/\npix}
  \draw[thick] (\xmin,\ytick) -- (\xmin-0.05,\ytick);
  \node[left] at (\xmin-0.05,\ytick) {\k};
}

\draw[thick, white] (0,1) -- (0,1.22);

\end{tikzpicture}
    \caption{The discrete~$n=40$ voxel formulation of the domain shown in Figure~\ref{Fig_illustration_in_and_out}, visualised as an image. Due to the~$n\times n$ shape of the discretised domain, there is an additional region~$I_\mathrm{ext}$ (grey) that falls outside of the circular printing region.}
    \label{Fig_strip}
\end{figure}

This work formulates illumination planning mathematically rigorously as a least-squares problem with non-negativity constraint to truly converge to solutions. The proposed framework, TVAM Adaptive Illumination Design (AID), is made using the Core Imaging Library~(CIL)~\cite{CILpaper,CILcode} to compare penalty functions and their effects on the energy dose profile thus allowing the solution to be tailored to selected printing metrics of interest such as in-part dose range. Moreover, the impact of the selection of design parameters ``upper and lower thresholds'' are investigated for both existing optimisation method OSMO as well as a novel penalty function implemented in TVAM AID. OSMO is selected for comparison rather than DrTVAM for using the same operation that maps an illumination plan onto an energy dose profile and for being an illumination planner often used in practice by the TVAM community~\cite{newGermanVAMGroup, newGermanVAMGroup2, edgeEnhanced, 3projectors}. The creation of TVAM AID and making use of CIL, as opposed to using an existing framework, is intentional. CIL is an open-source and maintained library for tomographic imaging with emphasis on reconstruction and is specifically chosen for reproducibility and extensibility. Its use lowers the barrier to entry as many operations are already implemented, enabling tools new to TVAM but established in classical CT reconstruction to be employed whilst also permitting a focus on the novel elements of penalty function design.

Thus, the contributions of this work can be summarised as:
\begin{enumerate}
    \item A critical analysis of penalty functions and their impact on selected printing metrics,
    \item Treatment of thresholds as design parameters with a systematic parameter sweep (on both the well-established OSMO method and the proposed novel framework) to characterise their effect on selected printing metrics and therefore aid in selecting thresholds for optimal print performance,
    \item A reproducible and readily extensible framework reducing the barrier to entry for novel illumination planning approaches.
\end{enumerate}

The remainder of this paper is organised as follows. In Section~\ref{section_methodology} (Methodology) the investigated penalty functions, or approaches, are outlined with Section~\ref{Section_method_metrics} detailing the metrics used to quantify their performance. The benefits, and drawbacks of the approaches are detailed in Sections~\ref{section_method_least_squares} to~\ref{section_method_mixed}. Results are presented and discussed in Section~\ref{section_resuts_and_discussion}, beginning with an investigation into the properties of the approaches so far included in TVAM AID in Section~\ref{section_comparison_proposed_methods}. The threshold parameter sweep is applied to both the best performing approach as well as OSMO in  Sections~\ref{section_thresholds} and~\ref{section_geometries}. Validation of the framework for a 3D geometry is presented in Section~\ref{section_3D}. Section~\ref{section_conclusions} presents conclusions and an outlook to future developments for this framework.

\section{Methodology}
\label{section_methodology}

For computation, the continuous domain~$\Omega$ is discretised into~$n \times n$ voxels with index~$i=1,2, \dots,n^2$. Thus,~$\Omegain$ is represented by the set~$I_\mathrm{in}$ of indices of voxels in the in-part region, and~$\Omegaout$ is represented by the set~$I_\mathrm{out}$ of indices of voxels in the out-of-part region. Hence, the printing domain where photopolymer is located is~$I_\mathrm{in} \cup I_\mathrm{out}$. Whilst the voxels are assembled in a quadratic image matrix, the cuvette is typically cylindrical with circular cross-section meaning there are voxels in the image that represent areas outside of the cuvette. In such cases, there is an additional region~$I_\mathrm{ext} \supset (I_\mathrm{in} \cup I_\mathrm{out})$ which should not be allowed to influence the illumination plan (Figure~\ref{Fig_strip}). Thus the energy dose image is represented by~$\textbf{f} \in \RR^{n^2}$, where~$\textbf{f}$ is a vertical stacking of all columns of voxel values. The voxel values of the energy dose image are naturally grouped into values associated with the `in', `out' and `external' domains:
\begin{enumerate}
    \item In-part voxels~$\textbf{f}_i~\textrm{for }i \in I_\mathrm{in}$ (hereafter denoted by~$\textbf{f}_\mathrm{in}$) that should receive sufficient light-energy to be cured,
    \item Out-of-part voxels~$\textbf{f}_i~\textrm{for }i \in I_\mathrm{out}$ (hereafter denoted by~$\textbf{f}_\mathrm{out}$) that should receive limited light-energy such that they remain uncured,
    \item Any remaining voxels~$\textbf{f}_i$ for~$i \in I_\mathrm{ext}$ may receive any light-energy dose with no penalisation for doing so since the region falls outside of the printing domain and thus where photopolymer that can absorb the light is present.
\end{enumerate}

In CT, an unknown image~$\textbf{f}$ is mapped by forward projection operator~$\textbf{A}$ to an observed set of projections, known as a sinogram~$\textbf{g}$. The aim, from the observed data~$\textbf{g}$ and knowledge of~$\textbf{A}$, is to reconstruct the image~$\textbf{f}$, i.e. to solve the linear system
\begin{equation}
    \textbf{Af} = \textbf{g}.
\end{equation}
The knowns and unknowns are reversed in TVAM. Given an illumination plan (sinogram)~$\textbf{g}$ that is back-projected by operator~$\AT$ (modelling the printing process), the achieved energy dose image
\begin{equation}
    \textbf{f} = \AT \textbf{g}
\end{equation}
is sought to be equal, or as close as possible, to~$\textbf{t}$ (the desired object to be printed). However, as mentioned in Section~\ref{section:introduction}, the non-negativity constraint on~$\textbf{g}$ typically requires an iterative approach. Instead, the problem is to find an optimal~$\textbf{g}^*$ that can reconstruct an~$\textbf{f}$ as close as possible to the target~$\textbf{t}$. Thus, finding a suitable TVAM illumination plan can be formulated as a least-squares problem with non-negativity constraint
\begin{equation} \label{eq:methods_optimisationGeneral}
    \textbf{g}^* = \argmin_{\textbf{g}} \| \AT \textbf{g} - \textbf{t} \|_2^2, \quad \textrm{s.t.~} \textbf{g} \geq \textbf{0}
\end{equation}
that seeks to minimise the sum of the squared residuals, i.e. the sum of the squares of the differences~$[\AT \textbf{g}]_i - \textbf{t}_i$ for~$i = {1, 2,\dots,n^2}$.

When using the binary material model,~$\textbf{t}$ is a binary image, i.e.
\begin{equation}
    \textbf{t}_i = 
    \begin{cases}
        0~\textrm{for }i \in I_\mathrm{out},\\
        1~\textrm{for }i \in I_\mathrm{in}
    \end{cases}
\end{equation}
where an energy value of~$0$ indicates liquid and an energy value of~$1$ indicates solid. Therefore, the general formulation~\eqref{eq:methods_optimisationGeneral} can be split into two separate parts to match the binary material model:
\begin{equation}
\begin{split}
    \| \AT \textbf{g} - \textbf{t} \|_2^2 & = \sum_{i=1}^{n^2} ([\AT \textbf{g}]_i - \textbf{t}_i)^2 \\
                                          & = \sum_{i \in I_{\mathrm{out}}} ([\AT \textbf{g}]_i - \textbf{t}_i)^2 + \sum_{i \in I_\mathrm{in}} ([\AT \textbf{g}]_i - \textbf{t}_i)^2 \\
                                          & = \sum_{i \in I_{\mathrm{out}}} p_\mathrm{out}([\AT \textbf{g}]_i) + \sum_{i \in I_\mathrm{in}} p_\mathrm{in}([\AT \textbf{g}]_i)
\end{split}
\end{equation}
where the two scalar functions~$p_\mathrm{out}$ and~$p_\mathrm{in}$ are used to encapsulate threshold knowledge. For example,
\begin{equation}
    p_\mathrm{out}(x) = (x - 0)^2 = x^2
\end{equation}
contains the information that voxels in~$I_\mathrm{out}$ have a target~$\textbf{t}_i = 0$ and
\begin{equation}
    p_\mathrm{in}(x) = (x - 1)^2
\end{equation}
that voxels in~$I_\mathrm{in}$ have a target~$\textbf{t}_i = 1$. Note that when~$I_\mathrm{ext} \neq \varnothing$, no penalties on any values in~$I_\mathrm{ext}$ occur for any and all approaches:
\begin{equation}
    p_\mathrm{ext}(x) = 0~\textrm{for }x \in I_\textrm{ext}.
\end{equation}

In summary, the resulting optimisation problem
\begin{equation}
    \textbf{g}^* = \argmin_{\textbf{g}} \sum_{i \in I_{\mathrm{out}}} p_\mathrm{out}([\AT \textbf{g}]_i) + \sum_{i \in I_\mathrm{in}} p_\mathrm{in}([\AT \textbf{g}]_i) ,~\textrm{s.t.~} \textbf{g} \geq \textbf{0}
\end{equation}
forms the core of the TVAM AID framework used in this work where the modularity stems from the freedom to design and combine different~$p_\mathrm{out}$ and~$p_\mathrm{in}$ to improve print performance. This framework will serve as the foundation to explore a variety of~$p_\textrm{in}$ and~$p_\textrm{out}$ penalty functions.

In order to solve the proposed optimisation problems (which will be convex) characterised by their penalty functions, the Fast Iterative Shrinkage-Thresholding Algorithm (FISTA)~\cite{FISTA} is used; specifically the implementation from CIL.

\subsection{Relevant Metrics}
\label{Section_method_metrics}
To assess the success of the different approaches, and provide a quantitative comparison with existing illumination planning, a variety of metrics can be employed. These metrics evaluate the suitability of an energy dose image~$\textbf{f}$. The three metrics listed below are adapted from the definitions of Rackson\etal~\cite{OSMO}.

\textbf{Process Window ($\PW$)} describes the gap between the minimal energy dose value observed by voxels~$\textbf{f}_\mathrm{in}$, and the maximum energy dose value observed by voxels~$\textbf{f}_\mathrm{out}$, i.e.
\begin{equation}
    \PW = \min(\textbf{f}_\mathrm{in}) - \max(\textbf{f}_\mathrm{out}).
    \label{eq:PW}
\end{equation}
This is adjusted from the definition by Rackson\etal~\cite{OSMO} given as~$\max(\mathbf{f}_\mathrm{out}) - \min(\mathbf{f}_\mathrm{in})$ which measures the overlap between the light-energy values of either region. Instead,~\eqref{eq:PW} measures the gap between the light-energy values of either region; with a negative value indicating no gap and thus a printed geometry that cannot match~$\textbf{t}$, and a positive value measuring the width of the gap. This change is done as it is the gap itself that is of interest. A large~$\PW$ is desirable since, in practice, this determines the difference in degree of cure between the solidified in-part and liquid out-of-part. This benefits process robustness by increasing flexibility in illumination stopping time as a small amount of extra illumination will be less likely to be sufficient to cause any voxel in ~$\textbf{f}_\mathrm{out}$ to be $ \geq \tau$, whether that be due to small variations in the chemistry of photopolymer or unknown printing time.

\textbf{In-Part Dose-Range ($\IPDR$)} quantifies the range of $\mathbf{f}_\mathrm{in}$ and is here defined as
\begin{equation}
    \IPDR = \max(\mathbf{f}_\mathrm{in}) - \min(\mathbf{f}_\mathrm{in}).
    \label{eq:IPDR}
\end{equation}
During printing, the illumination plan repeats every cuvette rotation. In the binary material model a voxel cures when its cumulative dose first exceeds threshold~$\tau$. Achieving this dosage may occur at different times for different voxels, which is governed by the illumination plan. Multiple rotations simply scale the dosage received. Hence, a smaller~$\IPDR$ concentrates the window in which cure times occur across the part and increases the likelihood that voxels cure within a narrow range of rotations (ideally within one rotation). Synchronising cure times is desirable because it shortens the interval in which solid and liquid coexist, reducing sedimentation of the denser cured polymer, and minimises the period during which light must pass through cured regions whose refractive index distorts the projected images. For these reasons, keeping~$\IPDR$ small is a key objective of an illumination plan. This definition of~$\IPDR$ is again adapted from the definition by Rackson\etal~\cite{OSMO} which defines $\IPDR$ as $1 - \min(\mathbf{f}_\mathrm{in})$ using normalised units of dose. As will be further discussed in Section~\ref{section_method_mixed}, the approaches investigated here to create illumination plans use absolute rather than relative values and thus there is no guarantee that $\max(\mathbf{f}_\mathrm{in}) = 1$; hence the adjustment.

\textbf{Voxel Error Rate ($\VER$)} is somewhat related to~$\PW$ in that it is a measure of the overlap `volume' $W$ between $\mathbf{f}_\mathrm{in}$ and $\mathbf{f}_\mathrm{out}$ (i.e. the count of voxels where $\mathbf{f}_\mathrm{out} > \min(\mathbf{f}_\mathrm{in})$) normalised by the total number of voxels~$n^2$ in $I_\mathrm{in} \cup I_\mathrm{out}$, i.e.
\begin{equation}
    \VER = \frac{W}{n^2}
    \label{eq:VER}
\end{equation}
which is the ratio of voxels in~$I_\mathrm{out}$ exceeding~$\min(\textbf{f}_\mathrm{in})$ to the total number of voxels in the printable region. A $\VER > 0$ means that~$\mathbf{f}$ cannot be segmented by any~$\tau$ and subsequently binarised such that it exactly matches~$\textbf{t}$. Thus, it is desired to obtain~$\VER = 0$. This definition remains unchanged from that given by Rackson\etal~\cite{OSMO}.

A hazard of the above metrics is their susceptibility to outliers; a single voxel with an erroneous dose (which may not be visible in histograms typically used to visualise dose distribution) could drastically change the value of any of these metrics. The proposed solution is to remove part of the data associated to outliers, i.e. given a percentage~$2\alpha$, in the definitions~\eqref{eq:PW},~\eqref{eq:IPDR} and~\eqref{eq:VER}, the maximums and minimums are replaced with~$(100-\alpha)\%$ and~$\alpha\%$ respectively. These adjusted metrics, denoted by superscript~$(100-\alpha)\%$ and subscript~$\alpha\%$ (e.g.,~$\PW_{\alpha\%}^{(100-\alpha)\%}$) and their significant difference when there are outliers in~$\textbf{f}$ will be explored in~Section~\ref{section_3D}.

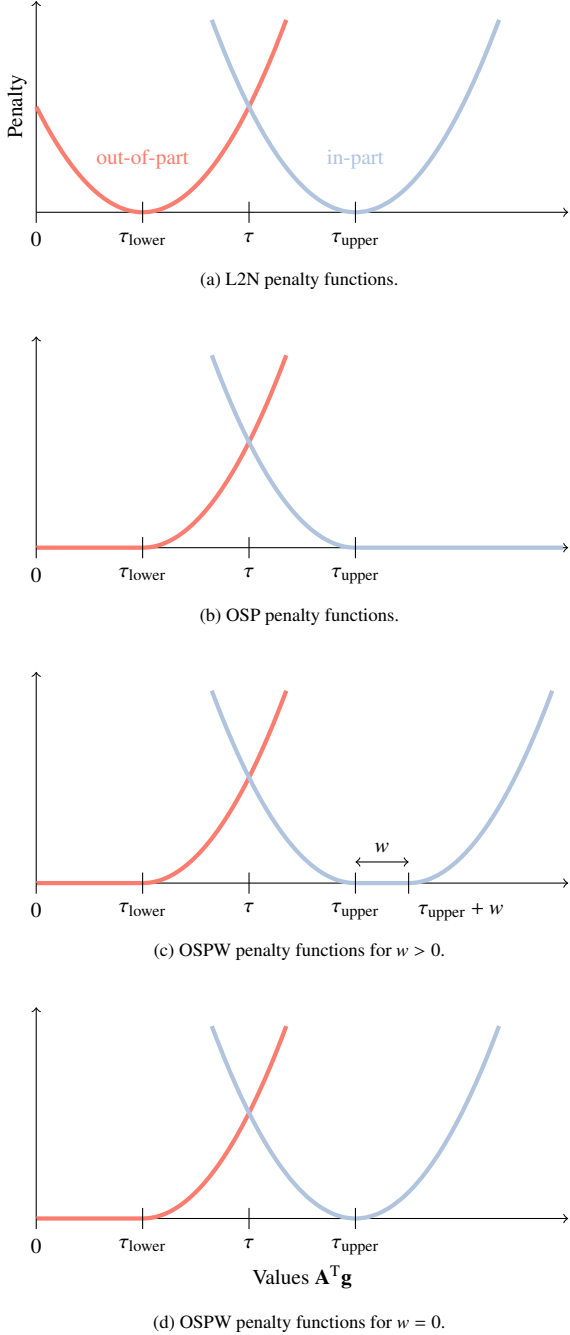
\begin{figure}
    \newcommand{\myshiftone}{1}
\newcommand{\myshifttwo}{3}
\newcommand{\myshiftthree}{3.5}
\newcommand{\myshift}{2}

\tikzset{
  font={\fontsize{8pt}{11}\selectfont}
}

\newcommand{\mylinewidth}{ultra thick}

\newcommand{\myscaling}{1.4}
\newcommand{\myfactor}{0.5}
\newcommand{\rightboundary}{5}
\newcommand{\upperboundary}{2}
\newcommand{\overallclip}{\clip (-0.2,-0.4) rectangle (\rightboundary+0.3,\upperboundary+0.4);}

\definecolor{myRed}{RGB}{250, 125, 111}
\definecolor{myBlue}{RGB}{176, 196, 222}

\begin{subfigure}{\linewidth}
    \centering
    \begin{tikzpicture}[scale=\myscaling]
        \draw (0,\upperboundary/2) node [rotate=90,above] {Penalty};
        \clip (-0.2,-0.4) rectangle (\rightboundary+0.3,\upperboundary+0);
    
        \draw (3,0.5) node[myBlue,font=\footnotesize] {in-part}; 
        \draw (1,0.5) node[myRed,font=\footnotesize] {out-of-part}; 
    
        \draw[->] (0,0) -- (\rightboundary,0);
        \draw[->] (0,0) -- (0,\upperboundary);
    
        \draw[\mylinewidth,scale=0.5, domain=-2:2.7, smooth, variable=\x, myRed] plot ({\x+2*\myshiftone}, {\myfactor*\x*\x});
        \draw[\mylinewidth,scale=0.5, domain=-2.7:2.7, smooth, variable=\x, myBlue] plot ({\x+2*\myshifttwo}, {\myfactor*\x*\x});
        
        \draw[] (0,0) -- (0,-0.1) node[below] {$0$};
        \draw[] (\myshiftone,0.1) -- (\myshiftone,-0.1) node[below] {$\taulower$};
        \draw[] (\myshifttwo,0.1) -- (\myshifttwo,-0.1) node[below] {$\tauupper$};
        \draw[] (\myshift,0.1) -- (\myshift,-0.1) node[below] {$\tau$};
    \end{tikzpicture}
    \caption{L2N penalty functions.}
\end{subfigure}

\begin{subfigure}{\linewidth}
    \centering
    \begin{tikzpicture}[scale=\myscaling]
        \draw (0,\upperboundary/2) node[rotate=90,above,white] {Penalty};
        \overallclip
    
    
        \draw[->] (0,0) -- (\rightboundary,0);
        \draw[->] (0,0) -- (0,\upperboundary);
    
        \draw[\mylinewidth,scale=0.5, domain=0:2.7, smooth, variable=\x, myRed] plot ({\x+2*\myshiftone}, {\myfactor*\x*\x});
        \draw[\mylinewidth,scale=0.5, domain=0:2*\myshiftone, smooth, variable=\x, myRed] plot ({\x}, {0});
        \draw[\mylinewidth,scale=0.5, domain=-2.7:0, smooth, variable=\x, myBlue] plot ({\x+2*\myshifttwo}, {\myfactor*\x*\x});
        \draw[\mylinewidth,scale=0.5, domain=2*\myshifttwo:\rightboundary*2-0.1, smooth, variable=\x, myBlue] plot ({\x}, {0});
    
        \draw[] (0,0) -- (0,-0.1) node[below] {$0$};
        \draw[] (\myshiftone,0.1) -- (\myshiftone,-0.1) node[below] {$\taulower$};
        \draw[] (\myshifttwo,0.1) -- (\myshifttwo,-0.1) node[below] {$\tauupper$};
        \draw[] (\myshift,0.1) -- (\myshift,-0.1) node[below] {$\tau$};
    \end{tikzpicture}
    \caption{OSP penalty functions.}
\end{subfigure}

\begin{subfigure}{\linewidth}
    \centering
    \begin{tikzpicture}[scale=\myscaling]
        \draw (0,\upperboundary/2) node[rotate=90,above,white] {Penalty};
        \overallclip
    
    
        \draw[->] (0,0) -- (\rightboundary,0);
        \draw[->] (0,0) -- (0,\upperboundary);
    
        \draw[\mylinewidth,scale=0.5, domain=0:2.7, smooth, variable=\x, myRed] plot ({\x+2*\myshiftone}, {\myfactor*\x*\x});
        \draw[\mylinewidth,scale=0.5, domain=0:2*\myshiftone, smooth, variable=\x, myRed] plot ({\x}, {0});
        \draw[\mylinewidth,scale=0.5, domain=-2.7:0, smooth, variable=\x, myBlue] plot ({\x+2*\myshifttwo}, {\myfactor*\x*\x});
        \draw[\mylinewidth, scale=0.5, domain=0:2.7, smooth, variable=\x, myBlue] plot ({\x+2*\myshiftthree}, {\myfactor*\x*\x});
        \draw[\mylinewidth,scale=0.5, domain=2*\myshifttwo:2*\myshiftthree, smooth, variable=\x, myBlue] plot ({\x}, {0});
    
        \draw[] (0,0) -- (0,-0.1) node[below] {$0$};
        \draw[] (\myshiftone,0.1) -- (\myshiftone,-0.1) node[below] {$\taulower$};
        \draw[] (\myshifttwo,0.1) -- (\myshifttwo,-0.1) node[below] {$\tauupper$};
        \draw[] (\myshift,0.1) -- (\myshift,-0.1) node[below] {$\tau$};
        \draw[] (\myshiftthree,0.1) -- (\myshiftthree,-0.1) node[below right] {$\tauupper + w$};
        \draw[<->] (\myshifttwo,0.2) -- (3.25,0.2) node[above] {$w$} -- (\myshiftthree,0.2);
    \end{tikzpicture}
    \caption{OSPW penalty functions for~$w > 0$.}
\end{subfigure}

\begin{subfigure}{\linewidth}
    \centering
    \begin{tikzpicture}[scale=\myscaling]
        \draw (0,\upperboundary/2) node[rotate=90,above,white] {Penalty};
        \draw (\rightboundary/2,0) node[rotate=0,below,yshift=-0.5cm] {Values $\AT \textbf{g}$};
        \overallclip
    
        
        \draw[->] (0,0) -- (\rightboundary,0);
        \draw[->] (0,0) -- (0,\upperboundary);
        
        \draw[\mylinewidth,scale=0.5, domain=0:2.7, smooth, variable=\x, myRed] plot ({\x+2*\myshiftone}, {\myfactor*\x*\x});
        \draw[\mylinewidth,scale=0.5, domain=0:2*\myshiftone, smooth, variable=\x, myRed] plot ({\x}, {0});
        \draw[\mylinewidth,scale=0.5, domain=-2.7:2.7, smooth, variable=\x, myBlue] plot ({\x+2*\myshifttwo}, {\myfactor*\x*\x});
    
        \draw[] (0,0) -- (0,-0.1) node[below] {$0$};
        \draw[] (\myshiftone,0.1) -- (\myshiftone,-0.1) node[below] {$\taulower$};
        \draw[] (\myshifttwo,0.1) -- (\myshifttwo,-0.1) node[below] {$\tauupper$};
        \draw[] (\myshift,0.1) -- (\myshift,-0.1) node[below] {$\tau$};
    \end{tikzpicture}
    \caption{OSPW penalty functions for~$w = 0$.}
\end{subfigure}
    \caption{Penalties on the out-of-part (red) and of the in-part (blue) for a) L2N which uses L2-norm penalties fitted to a specific target, b) OSP that uses one-sided penalties, c)~$\OSPW{>0.0}$ which uses one-sided penalties with a penalty on energy dose values greater than~$\tauupper + w$ to voxels in the in-part region, d)~$\OSPW{=0.0}$ is a special case of~$\OSPW{>0.0}$ which could be considered as using a one-sided penalty for voxels in the out-of-part region and an L2-norm penalty for voxels in the in-part region. An energy dose value achieved through an illumination plan will incur a penalty dependent on the specific value. In L2N, any deviation from the target energy values is punished with increasing cost, whereas in OSP, only deviation from below the target for the in-part region and above the target for the out-of-part region are punished. In~$\OSPW{>0.0}$, the penalties of OSP are used with an additional penalty if values exceed a given dose value which could be considered as a disjoint L2-norm penalty function with a no-penalty region of width~$w$. $\OSPW{=0.0}$ thus reduces the no-penalty region and therefore takes a combination of L2-norm penalties in in-part and OSP in out-of-part. The total cost of an illumination plan is the sum of the costs incurred by all voxels.}
    \label{fig:penalisation}
\end{figure}

\subsection{L2-norm penalties (L2N)}
\label{section_method_least_squares}

As mentioned in Section~\ref{section:introduction}, using rays of light to cure the photopolymer means that it is not possible to illuminate a region without also illuminating the regions behind or in front. So whilst it is possible to suppress the energy dose of a voxel to be below curing threshold, i.e.~$\textbf{f}_\mathrm{out} \leq \tau$, it is impossible to guarantee~$\textbf{f}_\mathrm{out} = \textbf{0}$. Similarly,~$\textbf{f}_\mathrm{in} = \textbf{1}$ is very difficult to achieve and is a stricter requirement than what is required for printing since~$\textbf{f}_\mathrm{in} > \tau$ is the `true' requirement in the binary material model. This informs the first exemplary approach which aims to consider threshold information in the optimisation process. As such, both penalty functions are defined as least squares functions
\begin{equation}
    p_\mathrm{out}(x) = (x - \taulower)^2
\end{equation}
and
\begin{equation} \label{eq:L2In}
    p_\mathrm{in}(x) = (x - \tauupper)^2
\end{equation}
where~$0\leq \taulower<\tau<\tauupper$ such that the penalties are shifted from the binary target to the thresholds (Figure~\ref{fig:penalisation}a). This approach will be referred to as L2N.

\subsection{One-sided penalties (OSP)}
\label{section_method_one_sided}

The L2-norm penalties that are used in L2N lack full knowledge of the binary material model and, thus, may be too strict. Specific to the binary material model used in this work, there is no need to penalise~$\textbf{f}_\mathrm{out} \leq \taulower$ and it is unnecessary to penalise~$\textbf{f}_\mathrm{in} \geq \tauupper$. Thus, L2N might not find printable solutions that do exist such as solutions that contain values~$\textbf{f}_\mathrm{out} \leq \taulower$ and/or~$\textbf{f}_\mathrm{in} \geq \tauupper$. Hence, the second approach, uses one-sided penalties
\begin{equation}
    p_\mathrm{out}(x) = 
    \begin{cases}
        0 &\textrm{for }x \leq \taulower,\\
        (x - \taulower)^2 &\textrm{for }x > \taulower
    \end{cases}
\end{equation}
and
\begin{equation}
    p_\mathrm{in}(x) = 
    \begin{cases}
        (x - \tauupper)^2 &\textrm{for }x < \tauupper,\\
        0 &\textrm{for }x \geq \tauupper,
    \end{cases}
\end{equation}
and will be referred to as OSP.

\subsection{Extension of OSP with Penalties on High Doses (OSPW)}
\label{section_method_mixed}
The one-sided~$p_\mathrm{in}$ that is used in OSP does not encourage a small~$\IPDR$ due the lack of penalty on high values in~$\textbf{f}_\textrm{in}$ as shown in Figure~\ref{fig:penalisation}b. As done by Wechsler\etal~\cite{waveOptical}, to encourage an upper limit to~$\textbf{f}$, the penalty for voxels~$\textbf{f}_\mathrm{in}$ is updated. The penalty for voxels~$\textbf{f}_\mathrm{out}$ remains unchanged as
\begin{equation} \label{eq:ospOut}
    p_\mathrm{out}(x) = 
    \begin{cases}
        0 &\textrm{for }x \leq \taulower,\\
        (x - \taulower)^2 &\textrm{for }x > \taulower
    \end{cases}
\end{equation}
and the updated penalty for voxels~$\textbf{f}_\mathrm{in}$ is
\begin{equation}
    p_\mathrm{in}(x) = 
    \begin{cases}
        (x - \tauupper)^2 &\textrm{for }x < \tauupper, \\
        0 &\textrm{for }\tauupper \leq x < \tauupper + w,\\
        (x - (\tauupper + w))^2 &\textrm{for }x \geq \tauupper + w
    \end{cases}
\end{equation}
that aims to restrict the maximum value of~$\textbf{f}_\mathrm{in}$ to~$\tauupper + w$ where~$w$ determines the width of the no penalty region. This approach will be referred to as~$\OSPW{>0.0}$.

The novelty in OSPW lies in the summation~$\tauupper + w$ not having a prescribed value, unlike in the work of Wechsler~\textit{et~al.} whose formulation defines~$\tauupper + w := 1$ to restrict the range of values so that~$\textbf{f} \in [0,1]$. This mimics the normalisation of~$\textbf{f}$ that occurs in many TVAM optimisation methods~\cite{OSMO, CAL, highFidelity} to force~$\textbf{f} \in [0,1]$. However, normalisation also scales the threshold values so they become relative rather than absolute. For example, if~$\tauupper = 0.9$ but the energy dose image is scaled by~$\max(\textbf{f})$, then~$\tauupper = 0.9/\max(\textbf{f})$ is also scaled and becomes defined relative to~$\max(\textbf{f})$. Thus, using a penalty over normalisation enables the thresholds to be defined in absolute terms without the loss of meaning that comes from scaling.

The parameter~$w$ has a clear connection to~$\IPDR$ since~$\textbf{f}_\textrm{in}$ is permitted to attain values between~$\tauupper$ and~$\tauupper + w$ without any further scrutiny. Therefore, there is a balancing act in reducing~$\IPDR$ by minimising~$w$ and keeping the penalties as relaxed as possible to encourage exploration of solutions. At the extreme,~$w = 0$ such that the penalty functions become
\begin{equation}
    p_\mathrm{out}(x) = 
    \begin{cases}
        0 &\textrm{for }x \leq \taulower,\\
        (x - \taulower)^2 &\textrm{for }x > \taulower
    \end{cases}
\end{equation}
and 
\begin{equation}
    p_\mathrm{in}(x) = (x - \tauupper)^2.
\end{equation}    
This novel approach,~$\OSPW{=0.0}$, is the final considered in this work, which can alternatively be considered a combination of OSP~\eqref{eq:ospOut} for~$\textbf{f}_\mathrm{out}$ and L2N~\eqref{eq:L2In} for~$\textbf{f}_\mathrm{in}$.

\section{Results and Discussion}
\label{section_resuts_and_discussion}

\begin{figure*}
    \centering
    \begin{subfigure}{\linewidth}
        \centering
        \includegraphics[]{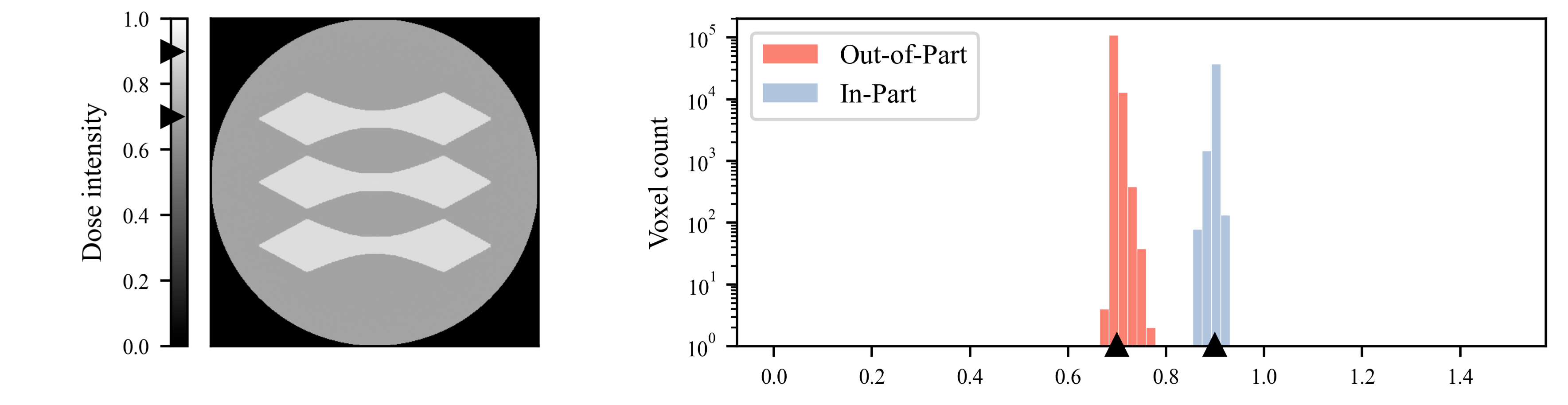}
        \caption{L2N derived result.}
    \end{subfigure}
    \par \bigskip
    \begin{subfigure}{\linewidth}
        \centering
        \includegraphics[]{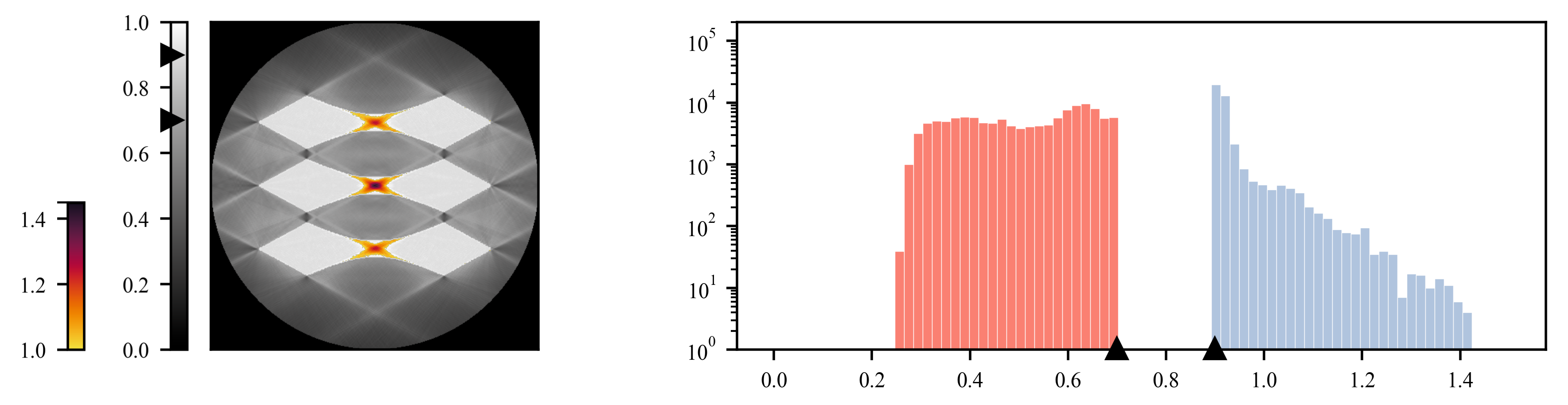}
        \caption{OSP derived result.}
    \end{subfigure}
    \par \bigskip
    \begin{subfigure}{\linewidth}
        \centering
        \includegraphics[]{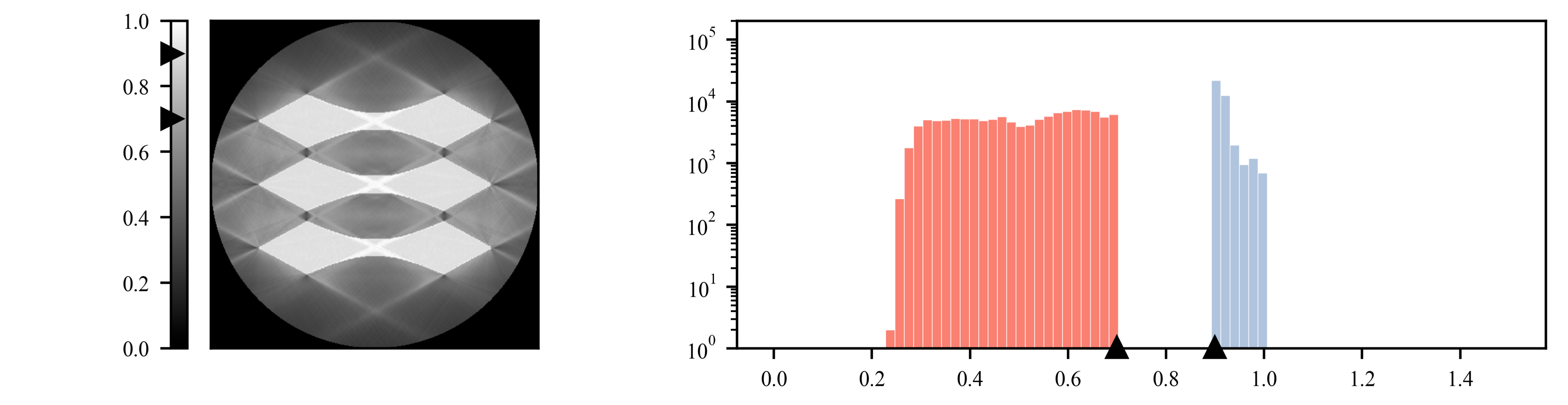}
        \caption{$\OSPW{=0.1}$ derived result.}
    \end{subfigure}
    \par \bigskip
    \begin{subfigure}{\linewidth}
        \centering
        \includegraphics[]{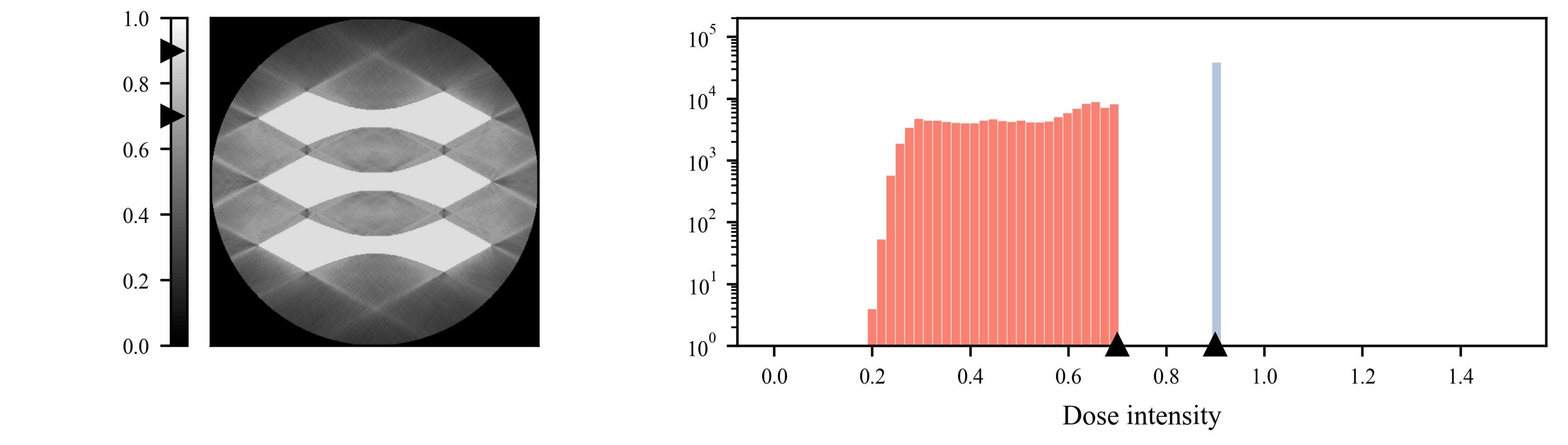}
        \caption{$\OSPW{=0.0}$ derived result.}
    \end{subfigure}
    \caption{A comparison of achieved dose profiles (right) and their respective histograms (left) created by each of the four approaches with 1000 iterations. Thresholds are fixed at~$\taulower = 0.70$ and~$\tauupper = 0.90$ and indicated by the black triangular markers.}
    \label{fig:exploringApproaches}
\end{figure*}
To begin, the four approaches are compared and characteristics of their achieved dose profiles discussed. Following observation of the~$\OSPW{=0.0}$ approach as producing the `best' achieved dose profile, the effects different combination pairs of~$\taulower$ and~$\tauupper$ have on the quality of the achieved dose profile are discussed. These results are compared to OSMO for a variety of geometries. Finally, the effectiveness of the methods on  a 3D geometry is presented, and again, contrasted to OSMO. Additionally, the susceptibility of the metrics to outlier values, as introduced in Section~\ref{Section_method_metrics}, is evaluated.

The code producing these results is available via the GitHub repository~\cite{projectGithub} and the code from~\cite{OSMOCode} was employed for an implementation of OSMO.

\begin{figure}
    \centering
    \includegraphics[width=\linewidth]{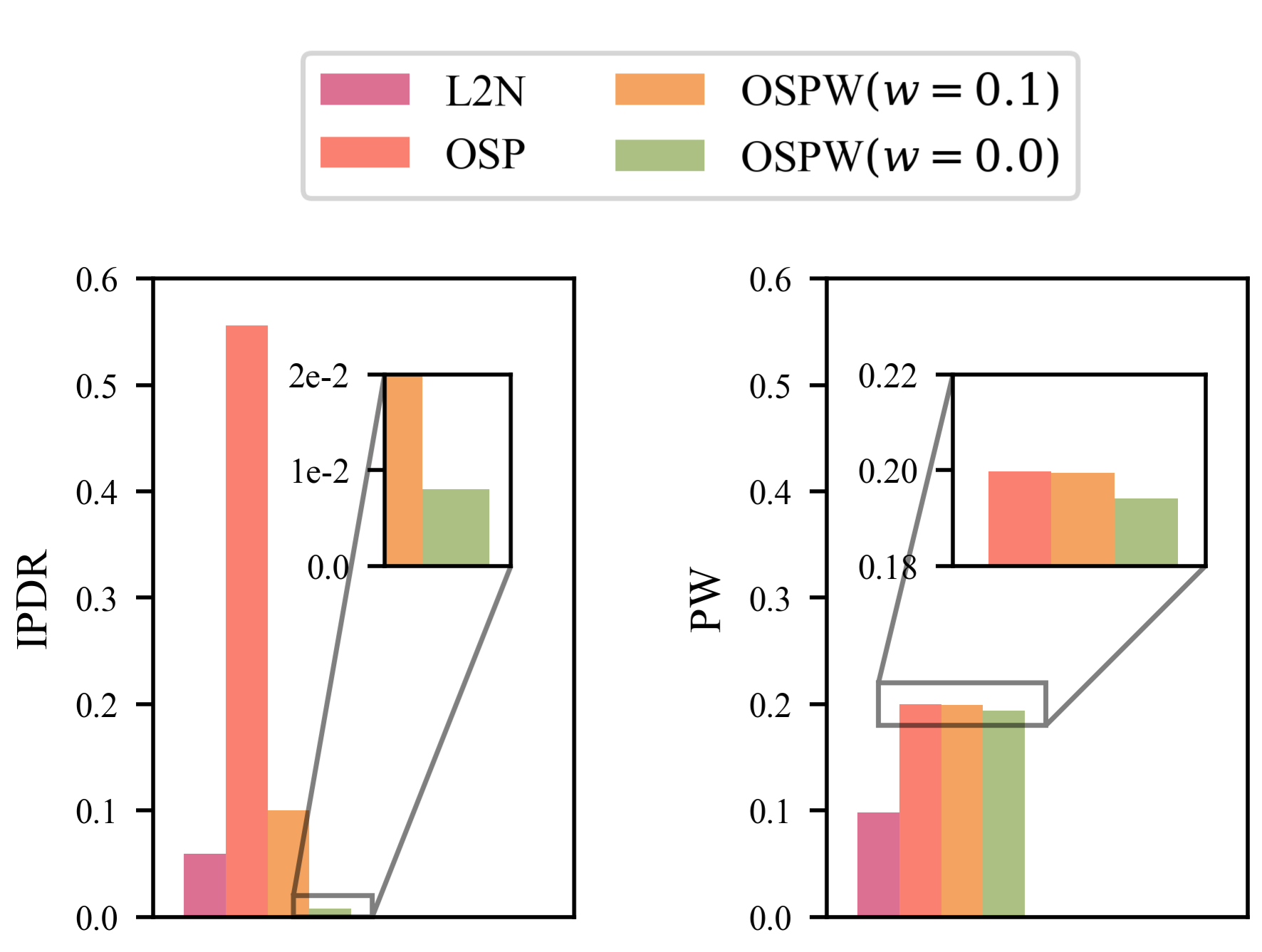}
    \caption{Summary of $\IPDR$ and $\PW$ for different approaches compared in Section~\ref{section_comparison_proposed_methods}.}
    \label{fig:exploringApproaches-metrics}
\end{figure}

\subsection{Comparison of Penalty Functions}
\label{section_comparison_proposed_methods}

A key feature of the TVAM AID framework is the flexibility of choice between different cost functions and constraints based on print requirements or desired characteristics. However, the choice comes with responsibility to select or design penalty functions that encourage certain behaviours in the energy dose profile. To demonstrate this, the four described approaches are applied to the DTU logo geometry which consists of three `dogbones' (in actuality representing lions~\cite{DTULogo}) containing hard corners and curved edges; see Figure~\ref{fig:exploringApproaches}a. Note that the image shown is achieved via L2N but is visually identical to the target geometry. In Figure~\ref{fig:exploringApproaches}, the achieved dose profiles obtained by said four approaches with prescribed thresholds~$\taulower = 0.70$ and~$\tauupper = 0.90$ are presented. Moreover, the~$\IPDR$ and~$\PW$ for each are plotted in Figure~\ref{fig:exploringApproaches-metrics}.

\textbf{L2N} shows an achieved dose profile that shows striking similarity to the target~(Figure~\ref{fig:exploringApproaches}a). This can be attributed to two observations. First, the appearance of uniform colour in the in-part and out-of-part regions which translates as near-uniform energy dose in each region. The histogram supports this by showing two narrow-width distributions driven by the parabolic shape of the L2-norm penalty functions applied to each region. Particular to the in-part histogram, the small width is associated with a small~$\IPDR$. Second, in the achieved dose profile, there is a stark difference between the achieved dose (indicated by highly different colourings) of the in-part and out-of-part regions leading to well-distinguished boundaries between the regions. In the histogram, this is shown by the separation of the two histograms i.e. a positive~$\PW$. However,~$\PW \neq \tauupper - \taulower$ as the optimisation does not result in perfect obedience to the upper and lower thresholds that are intended to be promoted via the L2-norm penalties. The peak of the in-part histogram is approximately located at~$\tauupper$ and the peak of the out-of-part histogram is approximately located at~$\taulower$. Consequently, it could be said that the~$\PW$ that could be achieved is eroded by approximately half the width of each histogram.

\begin{figure*}
    \centering
    \begin{subfigure}{\linewidth}
        \centering
        \includegraphics[]{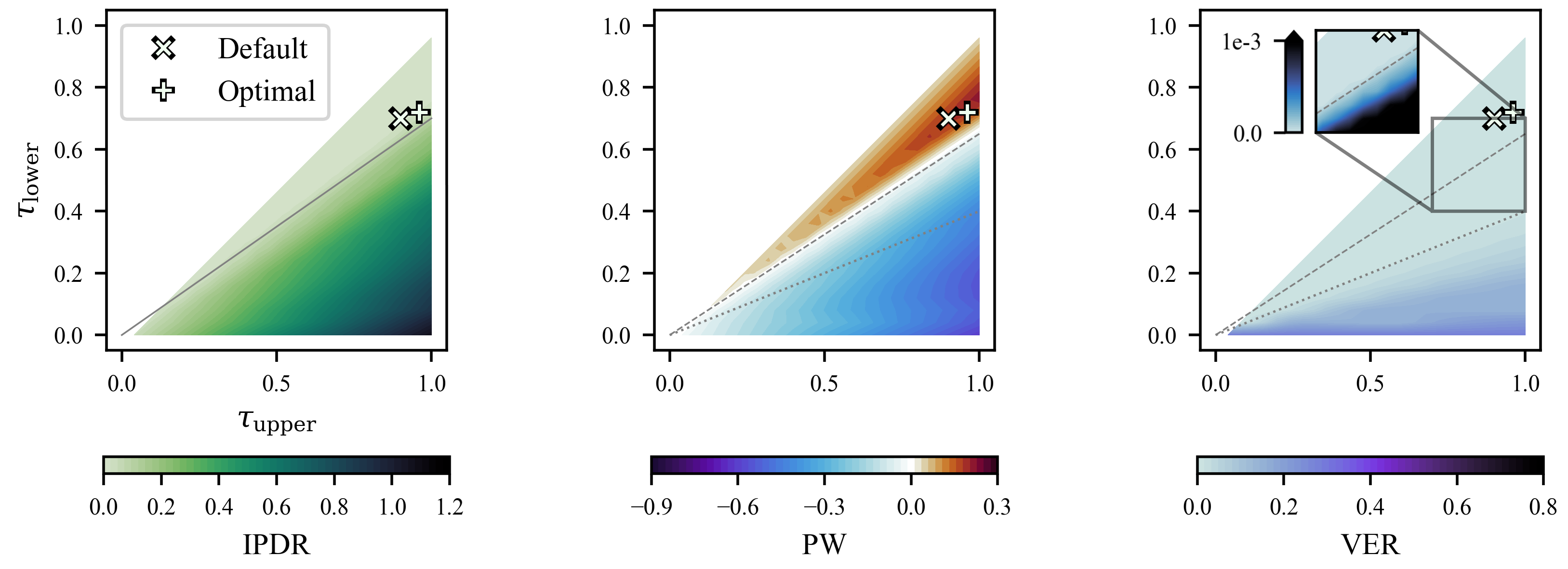}
        \caption{Colour maps for three critical metrics derived from the results of 325 threshold pairs using the~$\OSPW{=0.0}$ approach. The solid line~$\taulower = 0.75\tauupper$ (left) demarcates the edge of the region with near-constant~$\IPDR \approx 0$. The dashed line~$\taulower = 0.65\tauupper$ (centre, right) demarcates the edge of the region where~$\PW \geq 0$. The dotted line~$\taulower = 0.4\tauupper$ (centre, right) demarcates the apparent region where~$\VER=0$ but as the inset shows, this region is in fact bounded by~$\taulower = 0.65\tauupper$ which can only be seen by severely restricting the colour map.}
    \end{subfigure}
    \par \bigskip
    \begin{subfigure}{\linewidth}
        \centering
        \includegraphics[]{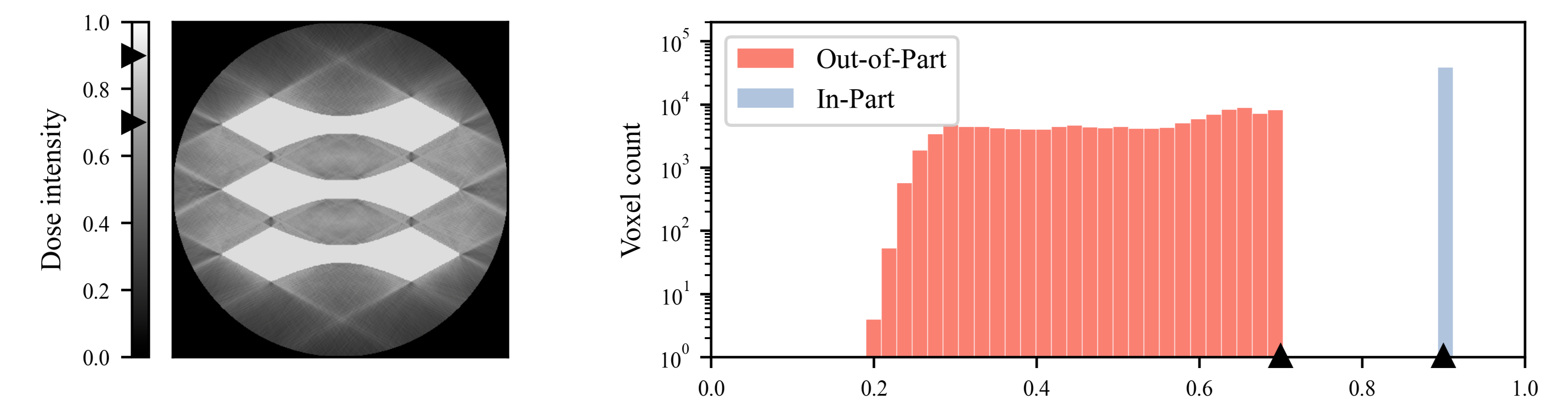}
        \caption{Achieved dose and corresponding histogram for the default threshold pair~$\taulower = 0.70$ and~$\tauupper = 0.90$ using the~$\OSPW{=0.0}$ approach.}
    \end{subfigure}
    \par \bigskip
    \begin{subfigure}{\linewidth}
        \centering
        \includegraphics[]{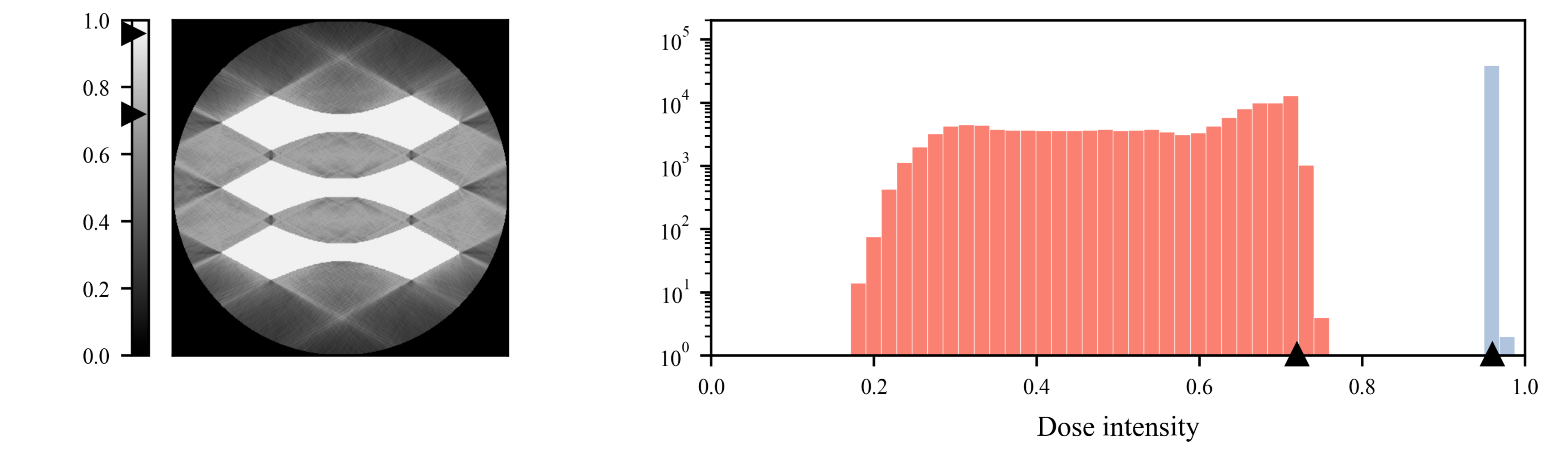}
        \caption{Achieved dose and corresponding histogram for the $\PW$-optimal threshold pair~$\taulower = 0.72$ and~$\tauupper = 0.96$ using the~$\OSPW{=0.0}$ approach.}
    \end{subfigure}
    \caption{The effects of threshold selection on the metrics attained after 1000 iterations of the~$\OSPW{=0.0}$ approach. The default and $\PW$-optimal threshold pairs lie close together so both provide similar achieved dose profiles with high~$\PW$ and small~$\IPDR$.}
    \label{fig:parameterChoiceMixed}
\end{figure*}

\begin{figure*}
    \centering
    \begin{subfigure}{\linewidth}
        \centering
        \includegraphics[]{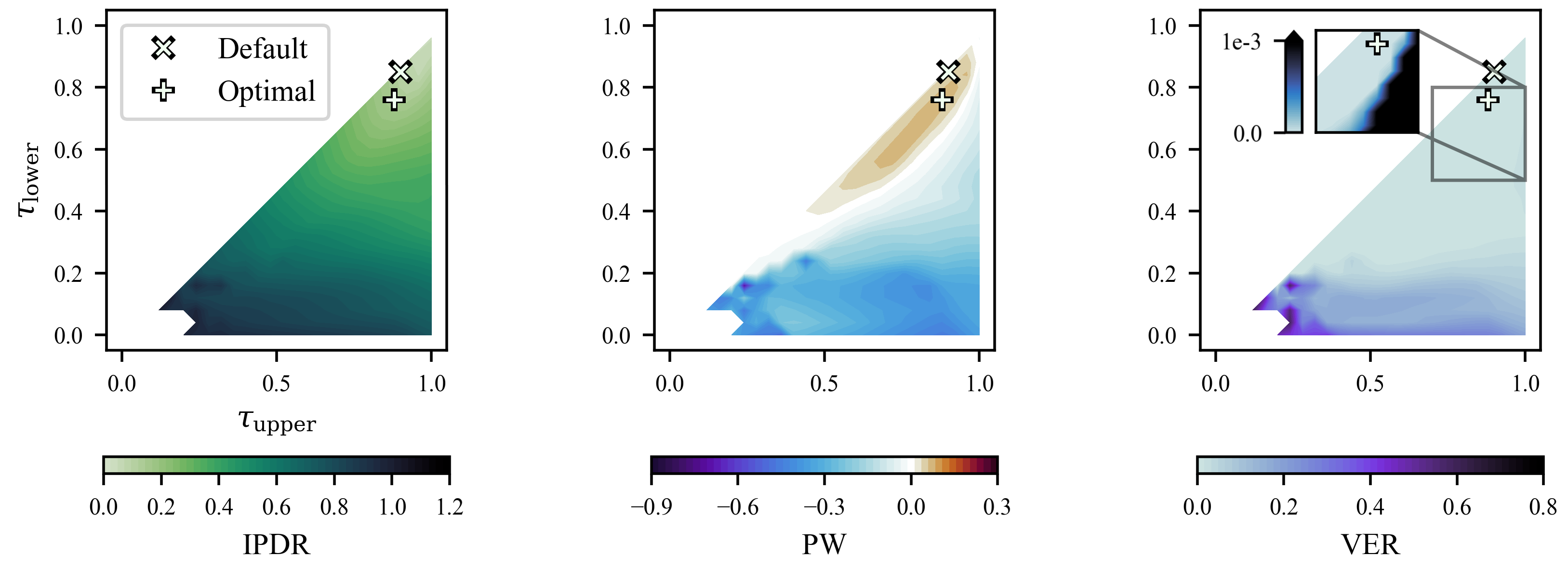}
        \caption{Colour maps for three critical metrics derived from the results of 325 threshold pairs using OSMO. Similarly to the~$\OSPW(=0.0)$ approach shown in Figure~\ref{fig:parameterChoiceMixed}, the true demarcation of the $\VER = 0$ region can only be seen with a restricted colour map.}
    \end{subfigure}
    \par \bigskip
    \begin{subfigure}{\linewidth}
        \centering
        \includegraphics[]{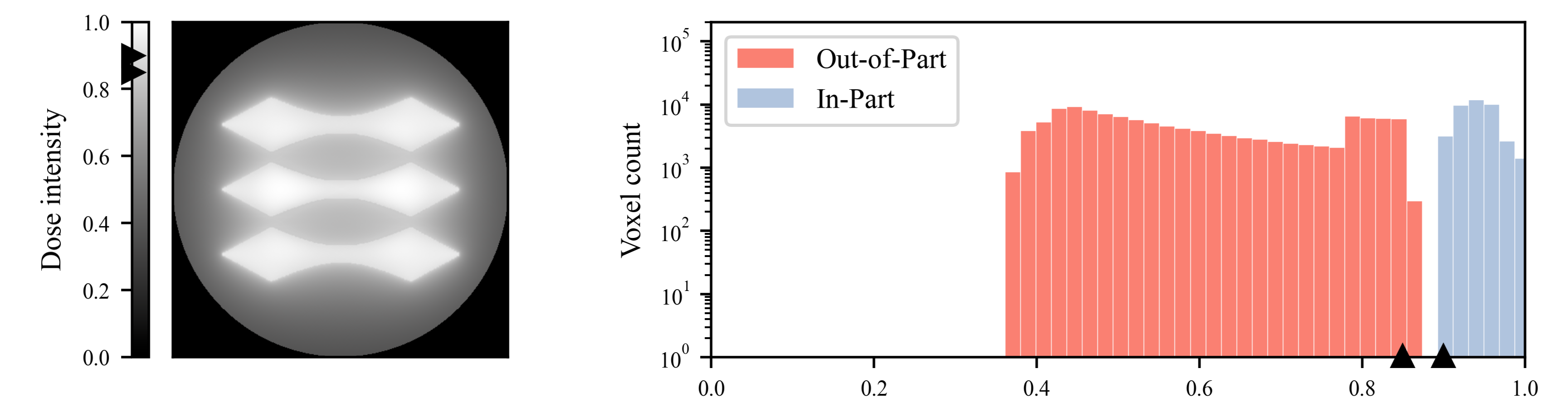}
        \caption{Achieved dose profile and corresponding histogram for the default threshold pair~$\taulower = 0.85$ and~$\tauupper = 0.90$ using OSMO.}
    \end{subfigure}
    \par \bigskip
    \begin{subfigure}{\linewidth}
        \centering
        \includegraphics[]{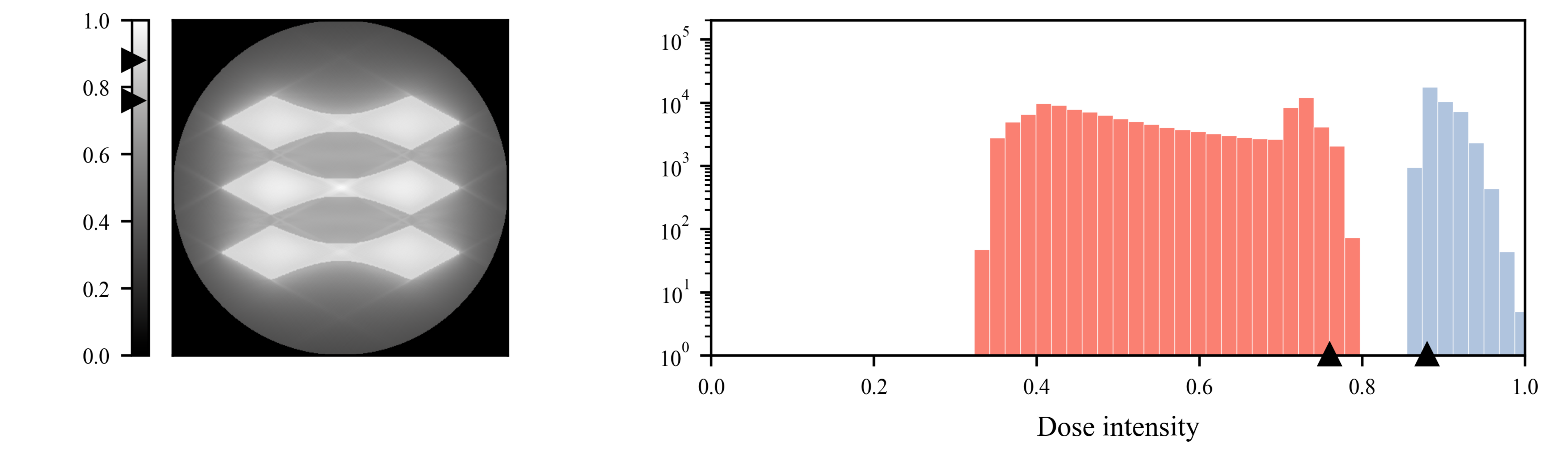}
        \caption{Achieved dose profile and corresponding histogram for the $\PW$-optimal threshold pair~$\taulower = 0.76$ and~$\tauupper = 0.88$ using OSMO.}
    \end{subfigure}
    \caption{The effects of threshold selection on the metrics attained after 1000 iterations of OSMO. The default and $\PW$-optimal threshold pairs lie close together but the increase in~$\PW$ this affords is significant. A note on the bottom left-hand corner of the OSMO metric colour maps where metrics are not calculated for these threshold pairs. It appears that the update to the out-of-part region of the so-called model results in a sinogram where all values are less than~$0$. As part of the processing of this sinogram, values below the so-called minimum projection value (defined as~$\geq 0$ and is set by default~$=0$) are re-valued to equal the minimum projection value. Thus, using default settings as is done here, the entire sinogram becomes a zero-value matrix and optimisation cannot continue constructively.}
    \label{fig:parameterChoiceOSMO}
\end{figure*}

\textbf{OSP} greatly increases the~$\PW$. The less restrictive one-sided penalties could be encouraging broader exploration during the optimisation which results in fewer voxels in~$I_\mathrm{in}$ achieving dose values less than~$\tauupper$ and fewer voxels in~$I_\mathrm{out}$ achieving dose values greater than~$\taulower$. Whilst this can be clearly seen in the larger separation of the two histograms~(Figure~\ref{fig:exploringApproaches}b), it is not immediately obvious in the achieved dose profile due to the non-uniform dose distribution. This is most widespread in the out-of-part region with many visible streaks. In the in-part region, large areas appear to continue to have near-uniform dose distribution but the thin central region of each `dogbone' is a source severe dose non-uniformity. This is due to the wider range of acceptable dose values in both regions allowed by the one-sided penalty contrary to L2N. The wide range of achieved dose values in the in-part region results in a very high~$\IPDR$ due to no upper constraint on dose values.

\textbf{OSPW}$\bm{(w=0.1)}$ successfully applies a penalty on voxels in~$I_\mathrm{in}$ that have a dose value greater than~$1$ leading to the in-part histogram~(Figure~\ref{fig:exploringApproaches}c) being less wide and, therefore, a marked improvement in~$\IPDR$. The additional penalty has very little effect on~$\PW$ and the general shape of the out-of-part histogram. Thus, in the achieved dose profile, the out-of-part region visually appears unchanged compared to the OSP result. The achieved dose profile for the in-part region also appears similar to OSP with the exception of the central regions; whilst they are still a source of dose non-uniformity, the magnitude of dose is now controlled. Outside of this desired improvement to the achieved in-part dose profile, the imperceptible changes elsewhere indicate that the penalties in OSP are too lax since they could still be adhered to here despite an additional penalty on high dose values.

\textbf{OSPW}$\bm{(w=0.0)}$ achieves a dose distribution~(Figure~\ref{fig:exploringApproaches}d) that highlights the flexibility of the TVAM AID framework to select penalty cost functions to achieve certain characteristics. The visually uniform dose distribution is in the in-part region where it matters; a characteristic possible due to the L2-Norm penalty and shown by the L2N result. In the out-of-part region, the more relaxed one-sided penalty is retained since non-uniform dose here is not a concern provided the dose values are below conversion threshold. Accordingly, as was seen with OSP and~$\OSPW{=0.1}$, the penalties of~$\OSPW{=0.0}$ that are more relaxed enable a far greater~$\PW$ to be achieved compared to L2N. Alternatively, the result here could be considered as retaining the large~$\PW$ achieved by OSP and~$\OSPW{=0.1}$ results whilst significantly decreasing~$\IPDR$ by `tightening' the penalty by using an L2-norm on voxels in~$I_\mathrm{in}$. The reduction is so great that~$\IPDR$ here is even lower than the L2N result which, like the increase in~$\PW$, could also be attributed to the relaxation of the penalty on voxels in~$I_\mathrm{out}$ which allows more freedom for the in-part L2-norm penalty to have a stronger effect.

Due to the combined characteristics of a wide~$\PW$ and a very small~$\IPDR$ leading to uniform dose distribution in the in-part region,~$\OSPW{=0.0}$ is considered to obtain the best performance compared to the other approaches (see Figure~\ref{fig:exploringApproaches-metrics}). Hence, this approach will be considered in more detail in the following sections. However, this is not to say that there is no merit in investigating the other approaches further, in particular, other choices of~$w \neq 0.0$ in OSPW. The freedom to select~$w$ is a key differentiator in the novel OSPW approach when compared to, for example, DrTVAM~\cite{inverseRendering, waveOptical} where~$w$ is not chosen and instead informed to ensure~$\tauupper + w \coloneq 1$. As seen in Figure~\ref{fig:exploringApproaches-metrics}, when~$w$ can be explicitly defined as~$w=0.0$, there is strong improvement in the performance of the achieved energy dose field suggesting OSPW is an improved penalty functional.

\subsection{Threshold Selection}
\label{section_thresholds}

Whilst the target geometry is a fixed parameter, there is the responsibility to choose threshold parameters~$\taulower$ and~$\tauupper$ which have a strong influence particularly on the obtained~$\PW$. If any optimisation adhered to the aforementioned penalties perfectly,~$\PW$ would simply be~$\tauupper - \taulower$. However, as previously discussed and shown in Figure~\ref{fig:exploringApproaches}, this is not necessarily the reality. For example, in the L2N result, the~$\PW$ that is achieved has been eroded by the positive width of each histogram. But, it may be possible that a different combination of thresholds with smaller~$\tauupper - \taulower$ could result in thinner histograms. Thus the erosion of~$\tauupper - \taulower$ may be lower leading to an overall wider~$\PW$. Conversely, it was shown that the other three approaches responded well to the tested thresholds and, in the case of~$\OSPW{=0.0}$, it would be of interest to further widen~$\PW$. However, there must be a limit since, as discussed in Section~\ref{section_method_least_squares}, the choice of $\taulower = 0$ and $\tauupper = 1$ will not yield satisfactory results due to the non-negativity constraint.

This motivates the following study to better understand the effect of threshold selection on the achieved dose profile and the consequences on their metrics. Threshold values are varied in the range~$(0,1)$ in increments of~$0.04$ such that there are 26 values each threshold can take. The resulting number of combinations (without repetitions thus naturally enforcing~$\taulower \neq \tauupper$) leads to~$\binom{26}{2} = 25(25+1)/2= 325$ pairs of thresholds. Since one value in the pair will always be strictly less than the other,~$\taulower < \tauupper$ can be easily enforced. As previously mentioned, results will focus on the~$\OSPW{=0.0}$ penalty approach. A fixed number of 1000 iterations is used by FISTA, found to satisfactorily ensure convergence of metrics for all threshold combinations.

The resulting colour maps for each of the three metrics are shown in Figure~\ref{fig:parameterChoiceMixed}a. The~$\IPDR$, which should be minimised, could generally be described as two attached planar surfaces approximately divided by the line~$\taulower = 0.75\tauupper$. The region~$\taulower \geq 0.75\tauupper$ is of interest as it has the lowest~$\IPDR$ and is effectively constant zero. The~$\PW$ colour map shows a more complex surface but has a clearly defined region~$\taulower \geq 0.65\tauupper$ where the metric is positive as desired. Similarly to~$\IPDR$, the colour map for~$\VER$ could also be described by two planar surfaces this time seemingly divided by~$\taulower = 0.4\tauupper$. In the same way, the upper region~$\taulower \geq 0.4\tauupper$ is of interest due to its seemingly constant~$\VER=0$. This is unexpected as a~$\PW \geq 0$ is required to obtain~$\VER = 0$. The inset figure shows part of the region~$0.4\tauupper < \taulower < 0.65\tauupper$ in greater detail with a smaller range scale. It can be seen that~$\VER \neq 0$, meaning there is an overlap between the two histograms. Further, the low magnitudes of~$\VER$ in this region indicate that only a small number of voxels are contained within this overlap. The fact that these small numbers of voxels lead to such a poor~$\PW$ indicates these erroneous voxels have significantly incorrect achieved dose values. This supports the notion of metric susceptibility to outliers introduced in Section~\ref{Section_method_metrics} which will be further investigated in Section~\ref{section_3D}.

\begin{figure*}
    \centering
    \begin{subfigure}{\linewidth}
        \centering
        \includegraphics[]{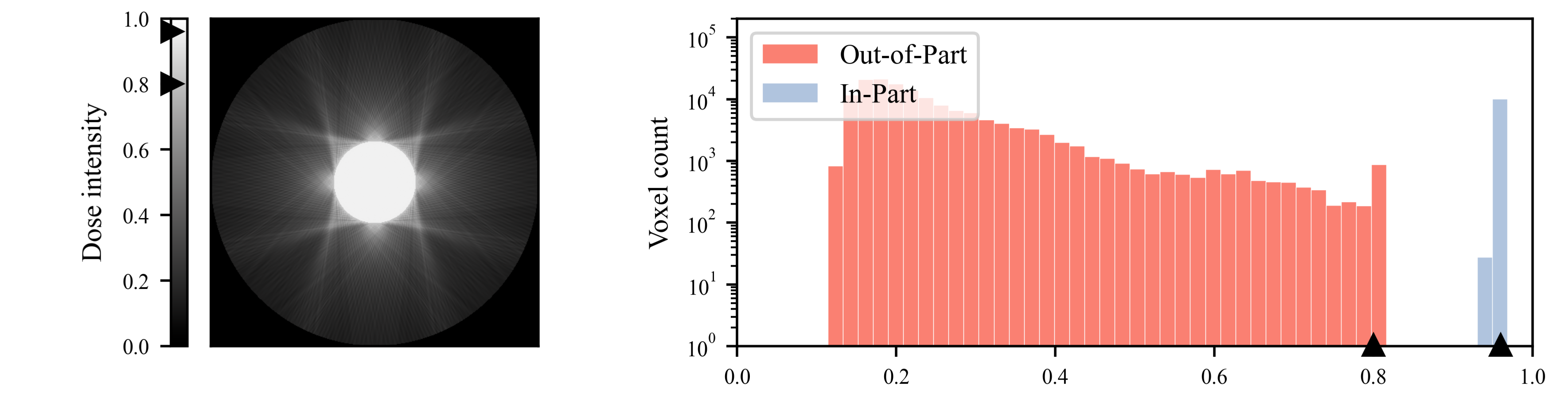}
        \caption{Achieved dose profile and corresponding histogram for the $\PW$-optimal threshold pair~$\taulower = 0.80$ and~$\tauupper = 0.96$ using~$\OSPW{=0.0}$.}
    \end{subfigure}
    \par \bigskip
    \begin{subfigure}{\linewidth}
        \centering
        \includegraphics[]{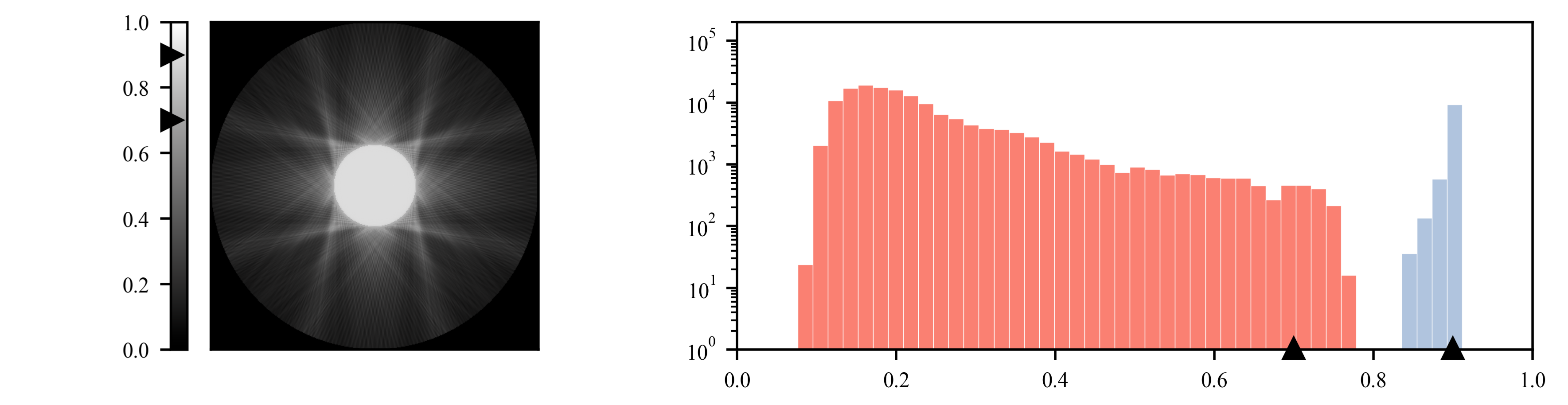}
        \caption{Achieved dose profile and corresponding histogram for the default threshold pair~$\taulower = 0.70$ and~$\tauupper = 0.90$ using~$\OSPW{=0.0}$.}
    \end{subfigure}
    \par \bigskip
    \begin{subfigure}{\linewidth}
        \centering
        \includegraphics[]{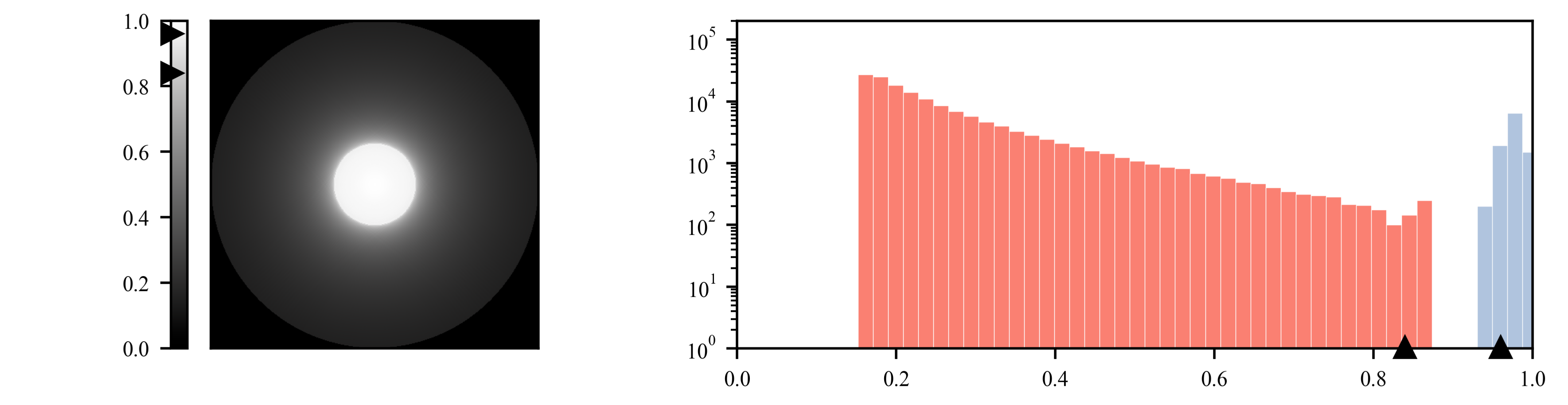}
        \caption{Achieved dose profile and corresponding histogram for the $\PW$-optimal threshold pair~$\taulower = 0.84$ and~$\tauupper = 0.96$ using OSMO.}
    \end{subfigure}
    \par \bigskip
    \begin{subfigure}{\linewidth}
        \centering
        \includegraphics[]{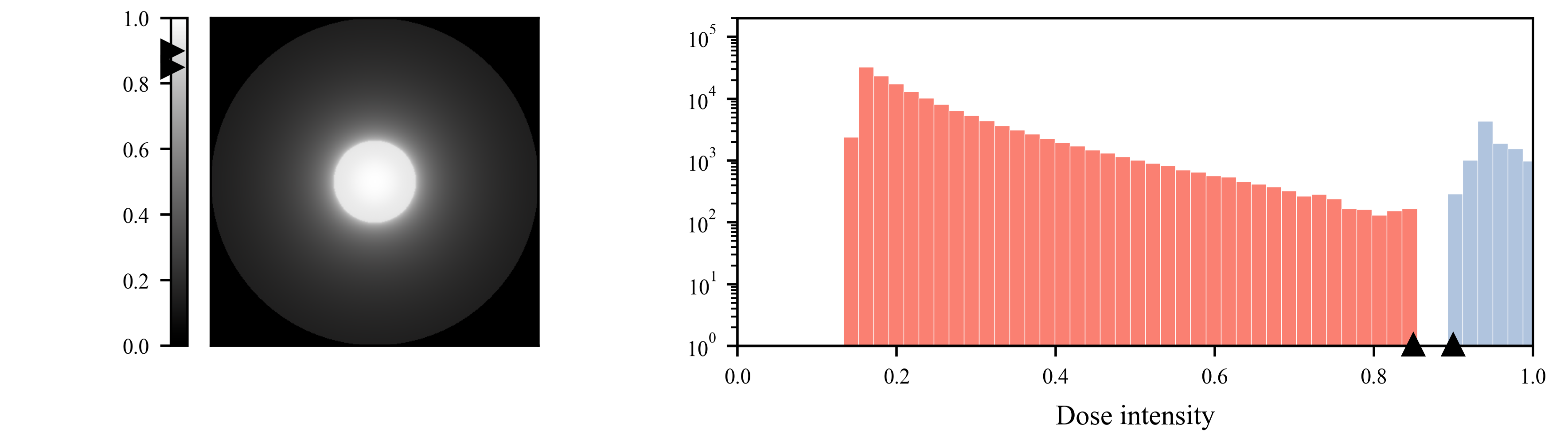}
        \caption{Achieved dose profile and corresponding histogram for the default threshold pair~$\taulower = 0.85$ and~$\tauupper = 0.90$ using OSMO.}
    \end{subfigure}
    \caption{Achieved dose profiles and corresponding histograms for a simple 2D disk after 1000 iterations. Rows are arranged:~$\OSPW{=0.0}$ $\PW$-optimal threshold pair,~$\OSPW{=0.0}$ default threshold pair, OSMO $\PW$-optimal threshold pair, OSMO default threshold pair.}
    \label{fig:geometryInfluenceDisk}
\end{figure*}

\begin{figure*}
    \centering
    \begin{subfigure}{\linewidth}
        \centering
        \includegraphics[]{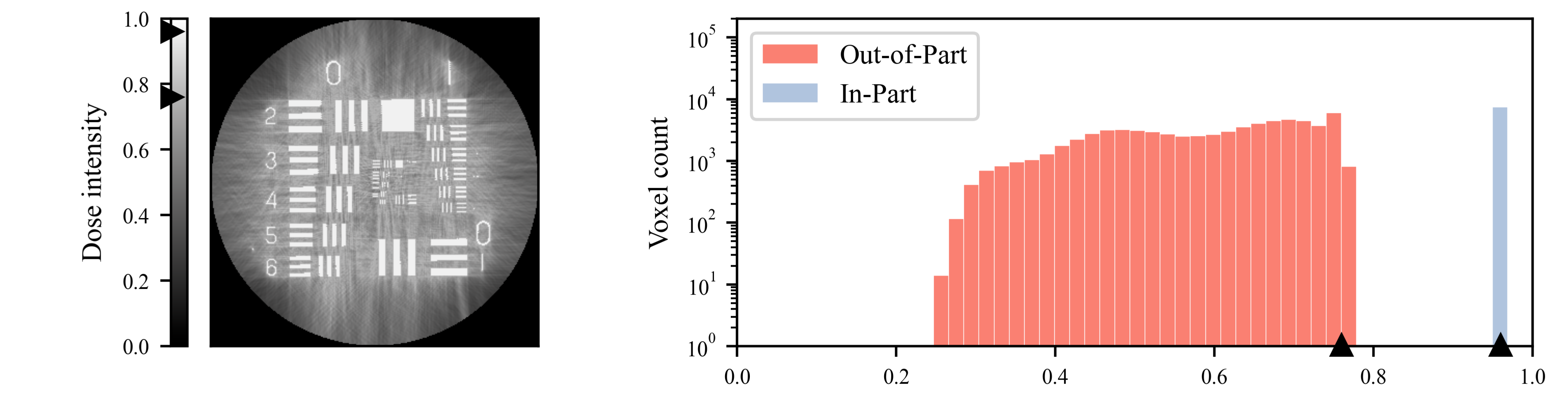}
        \caption{Achieved dose profile and corresponding histogram for the $\PW$-optimal threshold pair~$\taulower = 0.76$ and~$\tauupper = 0.96$ using~$\OSPW{=0.0}$.}
    \end{subfigure}
    \par \bigskip
    \begin{subfigure}{\linewidth}
        \centering
        \includegraphics[]{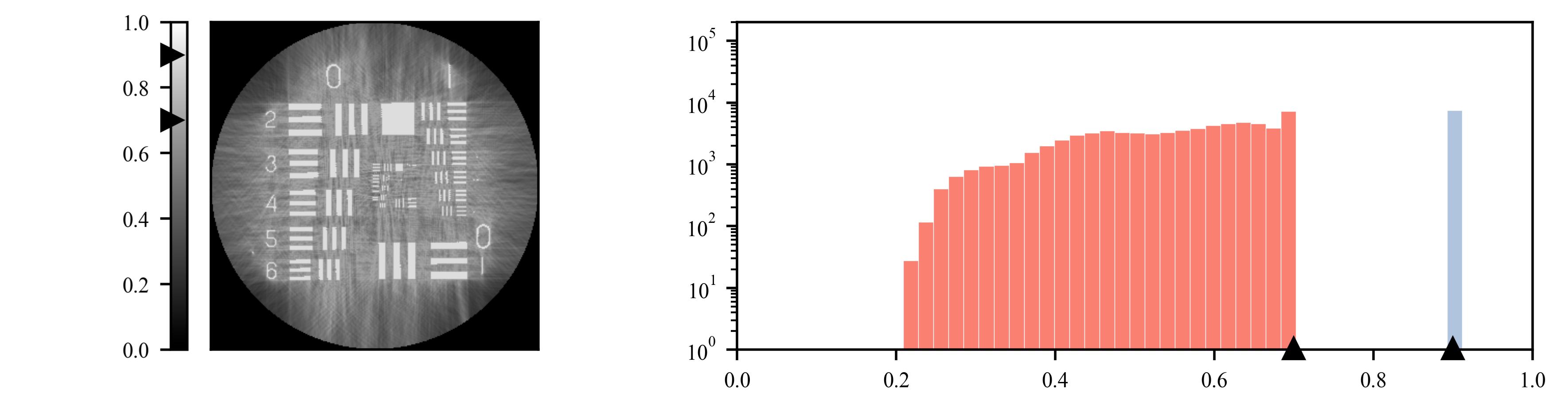}
        \caption{Achieved dose profile and corresponding histogram for the default threshold pair~$\taulower = 0.70$ and~$\tauupper = 0.90$ using~$\OSPW{=0.0}$.}
    \end{subfigure}
    \par \bigskip
    \begin{subfigure}{\linewidth}
        \centering
        \includegraphics[]{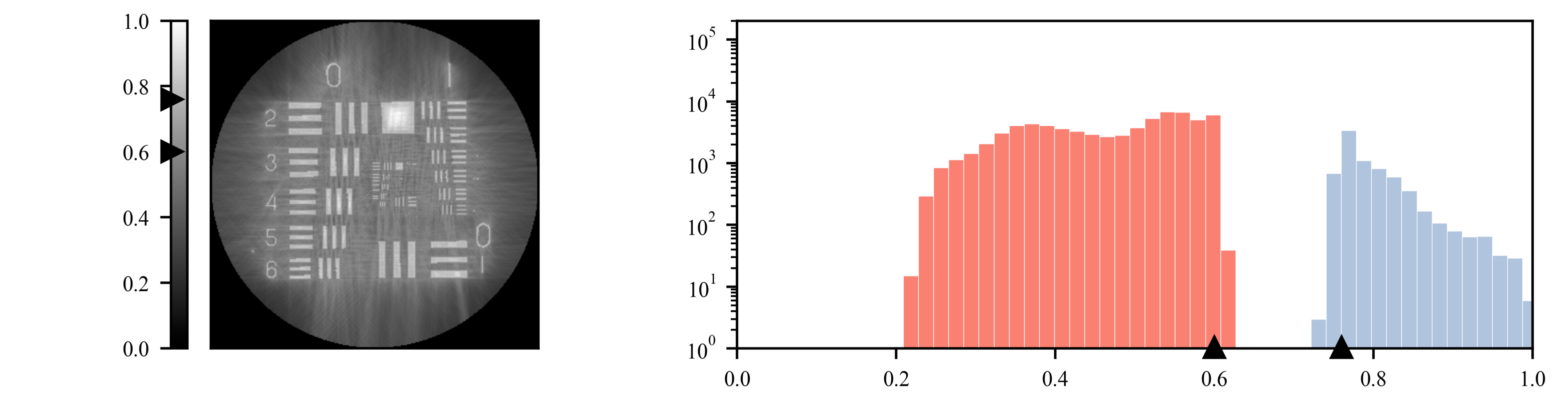}
        \caption{Achieved dose profile and corresponding histogram for the $\PW$-optimal threshold pair~$\taulower = 0.60$ and~$\tauupper = 0.76$ using OSMO.}
    \end{subfigure}
    \par \bigskip
    \begin{subfigure}{\linewidth}
        \centering
        \includegraphics[]{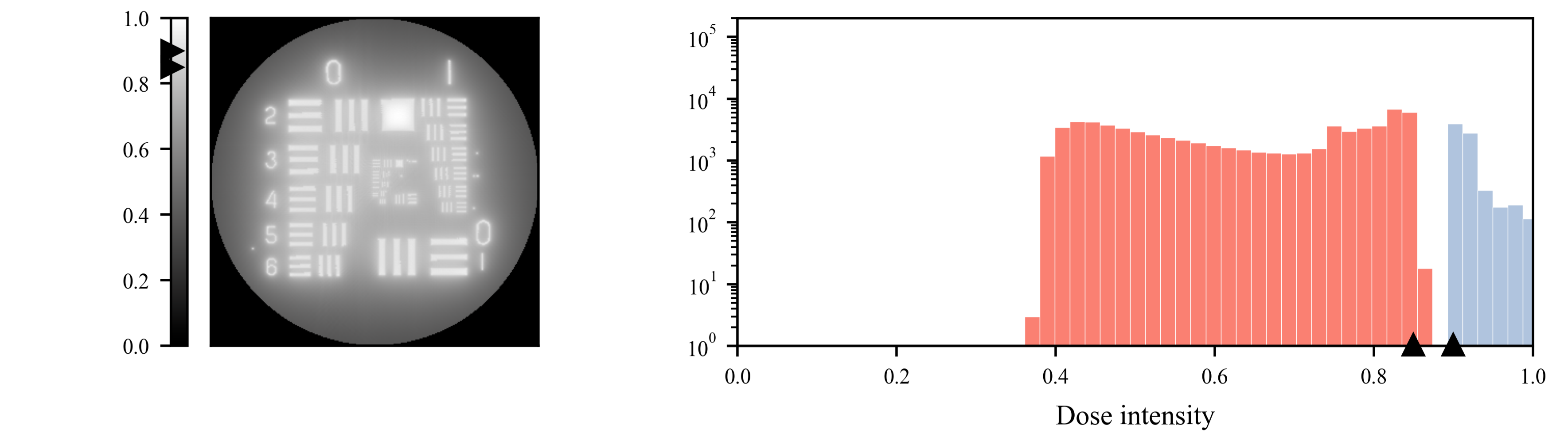}
        \caption{Achieved dose profile and corresponding histogram for the default threshold pair~$\taulower = 0.85$ and~$\tauupper = 0.90$ using OSMO.}
    \end{subfigure}
    \caption{Achieved dose profiles and corresponding histograms for a complex resolution test after 1000 iterations. Rows are arranged:~$\OSPW{=0.0}$ $\PW$-optimal threshold pair,~$\OSPW{=0.0}$ default threshold pair, OSMO $\PW$-optimal threshold pair, OSMO default threshold pair.}
    \label{fig:geometryInfluenceResolution}
\end{figure*}

\begin{figure*}
    \centering
    \includegraphics[width=\linewidth]{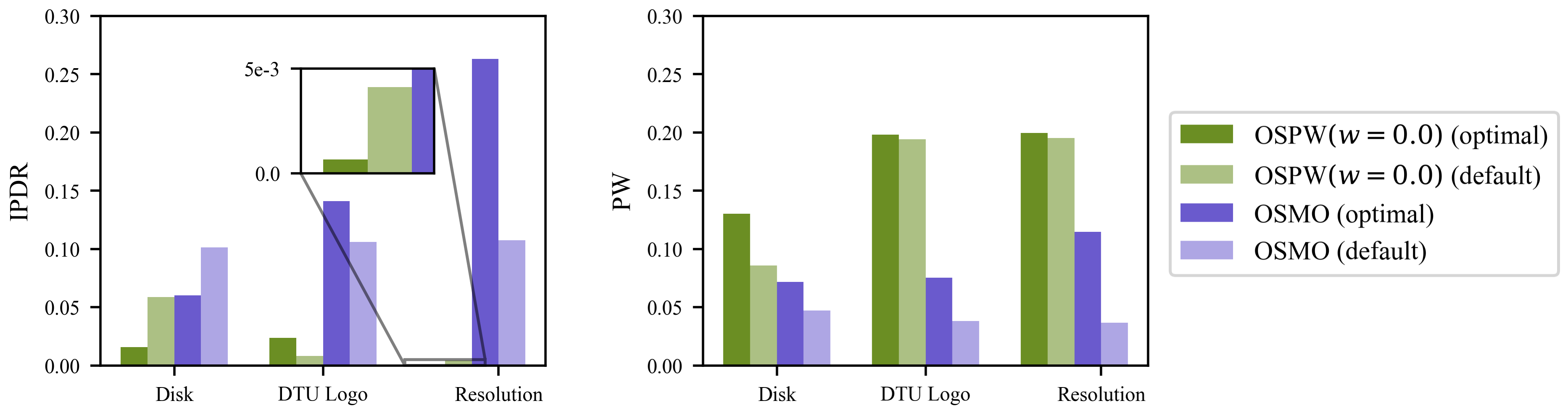}
    \caption{Summary of $\IPDR$ and $\PW$ for different test geometries used in Section~\ref{section_thresholds} and~\ref{section_geometries} (bar clusters). Both $\PW$-optimal and default thresholds of~$\OSPW{=0.0}$ and OSMO are used (bar colours).~$\OSPW{=0.0}$ consistently outperforms OSMO, even when comparing the results of~$\OSPW{=0.0}$ using default thresholds with OSMO using optimal thresholds.}
    \label{fig:geometryBars}
\end{figure*}

Since the desirable regions of the~$\IPDR$ and~$\VER$ colour maps greatly overlap and in these regions the metrics evaluate to a constant value (or near-constant in the case of~$\IPDR$), it stands to reason that any threshold pair in the region~$\taulower \geq 0.75\tauupper$ would adequately satisfy these two metrics. Thus, the search for the `optimal' threshold pair is now driven by~$\PW$ which does vary in this desired region. As mentioned,~$\PW \geq 0$ is the minimum requirement to achieve~$\VER=0$ which is necessary if an achieved dose profile is to be segmented by a single gelation threshold~$\tau$ and result in a perfect match to the target image. However, it has also been discussed that an increased~$\PW$ is sought for improvements to process robustness. Therefore, the `optimal' threshold pair is one that maximises~$\PW$ which, for this target image, is found to be~$\taulower = 0.72$ and~$\tauupper = 0.96$. Note that threshold pairs which result in a maximum achieved dose intensity greater than one are excluded from the selection process. Similar colour maps emerge for the other three approaches presented (Figure~\ref{fig:dogbonesHeatMaps}) as well as other target images which will be considered further in Section~\ref{section_geometries} (see also Figure~\ref{fig:diskHeatMaps} and Figure~\ref{fig:resolutionHeatMaps}). Therefore, the default threshold values recommended are~$\taulower = 0.70$ and~$\tauupper = 0.90$.

The achieved dose profile from the default threshold pair is shown in Figure~\ref{fig:parameterChoiceMixed}b. Compared to the achieved dose profile from the optimal threshold pair for this case in Figure~\ref{fig:parameterChoiceMixed}c, the most notable improvement is the brightness of the in-part region. This is expected, and supported by the histogram, due to the increase in~$\tauupper$ from 0.90 to 0.96 which shifts the entire in-part histogram in the positive direction with a small increase~$\IPDR$. This is offset by a slight increase in~$\PW$ as expected by the search for the best thresholds is solely driven by this metric. Overall, the differences between the two achieved dose profiles are minor, further reinforcing the appropriateness of the suggested default thresholds.

This general parameter sweep methodology is, of course, agnostic to the illumination planning technique used and so, as previously mentioned, the OSMO method is chosen to provide a comparison. In the original work~\cite{OSMO},~$\taulower = 0.85$ and~$\tauupper = 0.90$ are proposed as suitable defaults which are used here. For fairness, the $\PW$-optimal pair of thresholds is also found using the same parameter sweep method as above which yields~$\taulower = 0.76$ and~$\tauupper = 0.88$.

The OSMO colour map for~$\IPDR$~(Figure~\ref{fig:parameterChoiceOSMO}) could be described by isolines in approximately parabolic shape where the closer the threshold pair to the top right-hand corner, the better the~$\IPDR$. In contrast, the~$\PW$ colour map features a long thin valley of high~$\PW$. Similar to the~$\OSPW{=0.0}$~$\VER$ colour map, the OSMO~$\VER$ colour map appears to have a very large region where~$\VER = 0$ but as the inset shows,~$\VER \neq 0$ although the magnitudes are very low and the boundary of the region with good~$\VER$ score aligns with the~$\PW=0$ line.

The region of best achievable~$\IPDR$ does not have a large overlap with the region of best achievable~$\PW$. Fortunately, the $\PW$-optimal pair of thresholds, which results in a substantial increase in~$\PW$ compared to the default threshold~$\PW$, does not significantly compromise~$\IPDR$ which rises by a modest amount. However, the effects of this increase in~$\IPDR$ are seen in the optimal achieved dose profile where the in-part region is visibly non-uniform compared to the default achieved dose profile which appears more uniform. Concurrently, the increase in~$\PW$ reduces some of the blur at the boundaries between out-of-part and in-part regions that are seen in the default achieved dose profile. However, these improvements in~$\PW$ and~$\IPDR$ are not sufficient for the optimal thresholds OSMO result to surpass either the default or optimal thresholds~$\OSPW{=0.0}$ results for this target geometry; in both~$\OSPW{=0.0}$ cases~$\PW$ and~$\IPDR$ score far better~(Figure~\ref{fig:geometryBars}).

\subsection{Performance of Different Geometries}
\label{section_geometries}

Clearly, the biggest change that can be made to the optimisation is the geometry. Two further geometries are used to test~$\OSPW{=0.0}$. If the DTU logo of previous examples is considered a geometry of moderate complexity, it is reasonable to test on a very simple geometry of a centred disk (or cylinder in 3D) and a very complex geometry of a resolution test (included in the CIL data~\cite{CILcode}) designed to be intricate and difficult for which to obtain a suitable sinogram. The same method to find $\PW$-optimal thresholds as in Section~\ref{section_thresholds} is used here.

\textbf{Centred disk:} the two sets of~$\OSPW{=0.0}$ results in Figure~\ref{fig:geometryInfluenceDisk}a and~b illustrate what was mentioned at the start of Section~\ref{section_thresholds}: a larger~$\tauupper - \taulower$ does not always result in a larger~$\PW$. For this geometry, the default thresholds are quite poorly adhered to by the~$\OSPW{=0.0}$ achieved dose profile and, as the histogram shows, there are many voxels in~$I_\mathrm{out}$ that have a dose greater than~$\taulower$ as well as many voxels in~$I_\mathrm{in}$ that have a dose less than~$\tauupper$. In contrast, the $\PW$-optimal threshold pair produces an achieved dose profile with better respect for the thresholds so despite~$\tauupper - \taulower = 0.16$ (i.e. $0.04$ smaller than the difference of the default thresholds), there is a significant increase in~$\PW$. These less punishing penalties also serve to significantly decrease the~$\IPDR$ leading to a far superior achieved dose profile. As with the previous DTU logo geometry, an adequate increase in~$\PW$ is seen when comparing the OSMO dose profile achieved with default thresholds and the $\PW$-optimal thresholds dose profile~(Figure~\ref{fig:geometryInfluenceDisk}c and~d). For this geometry, however, the~$\IPDR$ also improves with the optimal thresholds. This makes the OMSO dose profile achieved with its optimal thresholds comparable, in terms of~$\PW$ and~$\IPDR$ values, to the~$\OSPW{=0.0}$ dose achieved with default thresholds. Perhaps most notable is that the~$\OSPW{=0.0}$ dose achieved with default thresholds outperforms the OSMO dose achieved with optimal thresholds for both metrics despite both sets of optimal thresholds (having similar interpretations as parameters) being almost identical.

\textbf{Resolution test:}~$\OSPW{=0.0}$ performs exceptionally; both default and optimal thresholds resulting in achieved dose profiles that have a wide~$\PW$ and very narrow~$\IPDR$~(Figure~\ref{fig:geometryInfluenceResolution}a and~b). Use of the $\PW$-optimal threshold pair for OSMO once again provides a substantial increase in~$\PW$ when comparing with the OSMO achieved dose profile from the default threshold pair~(Figure~\ref{fig:geometryInfluenceResolution}c and~d). The~$\IPDR$ increases significantly with use of the OSMO optimal threshold pair such that dose non-uniformity can be easily seen in the large square (right of centre) in achieved dose profile; for this geometry, a pair of thresholds for OSMO needs to compromise between the two metrics.

Taking an overview of~$\IPDR$ and~$\PW$ for all three geometries, both~$\OSPW{=0.0}$ and OSMO, and both default and optimal threshold pairs, Figure~\ref{fig:geometryBars} shows a clear outcome: when considering~$\IPDR$ which must be minimised,~$\OSPW{=0.0}$ outperforms OSMO for every geometry. There is additionally a trend that the more complex geometries show a greater improvement in metrics by using~$\OSPW{=0.0}$ over OSMO. This holds even when comparing~$\OSPW{=0.0}$ with default thresholds to OSMO with optimal thresholds. The same can be said of~$\PW$, which should be maximised. Furthermore, generally the difference in~$\PW$ and~$\IPDR$ for~$\OSPW{=0.0}$ with default thresholds and~$\OSPW{=0.0}$ with optimal thresholds is small further validating the choice of default thresholds.

\subsection{Validation for a 3D Geometry}
\label{section_3D}

\begin{figure}
    \centering
    \includegraphics[width=\linewidth]{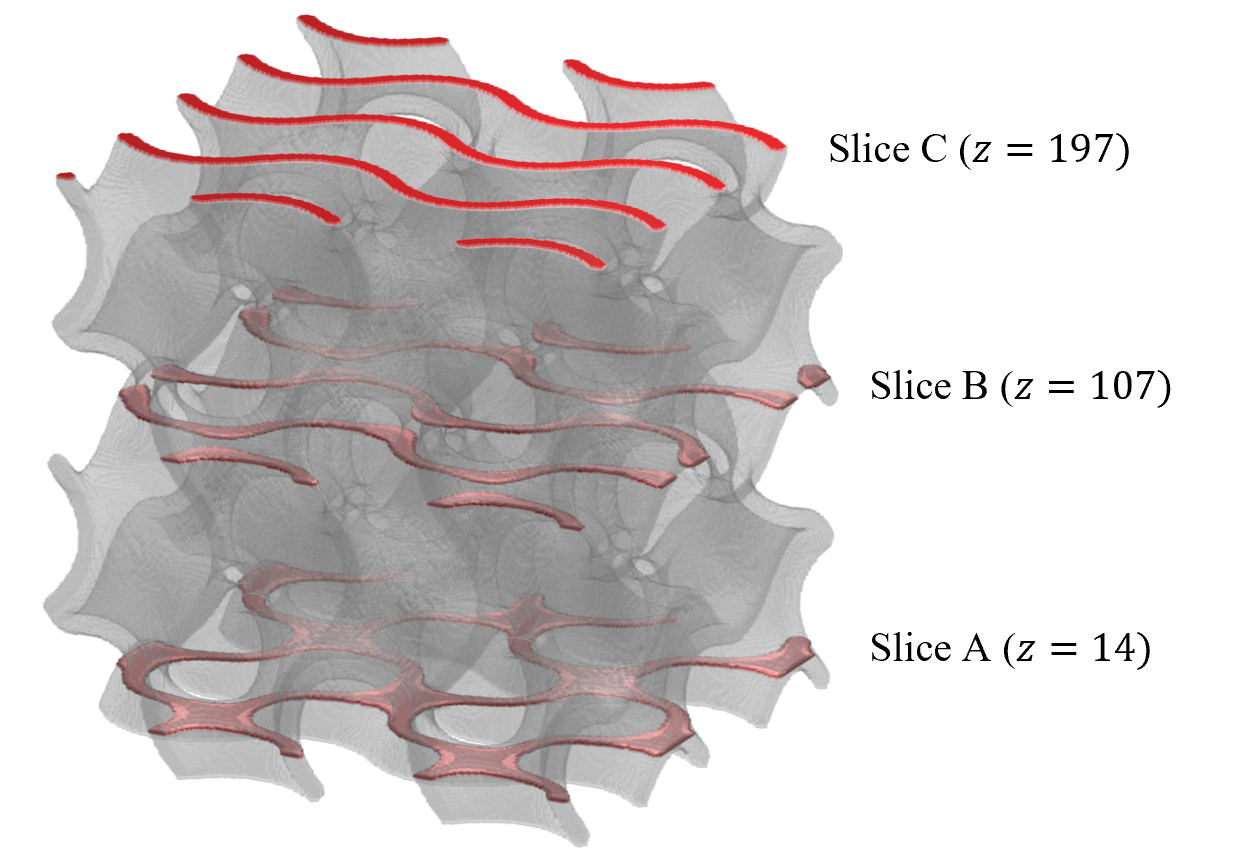}
    \caption{Gyroid 3D test geometry with eight unit cells ($2 \times 2 \times 2$). Reported visualisation slices at $z=[14,107,197]$ are highlighted.}
    \label{fig:geometry3D}
\end{figure}

\begin{figure*}
    \centering
    \includegraphics[]{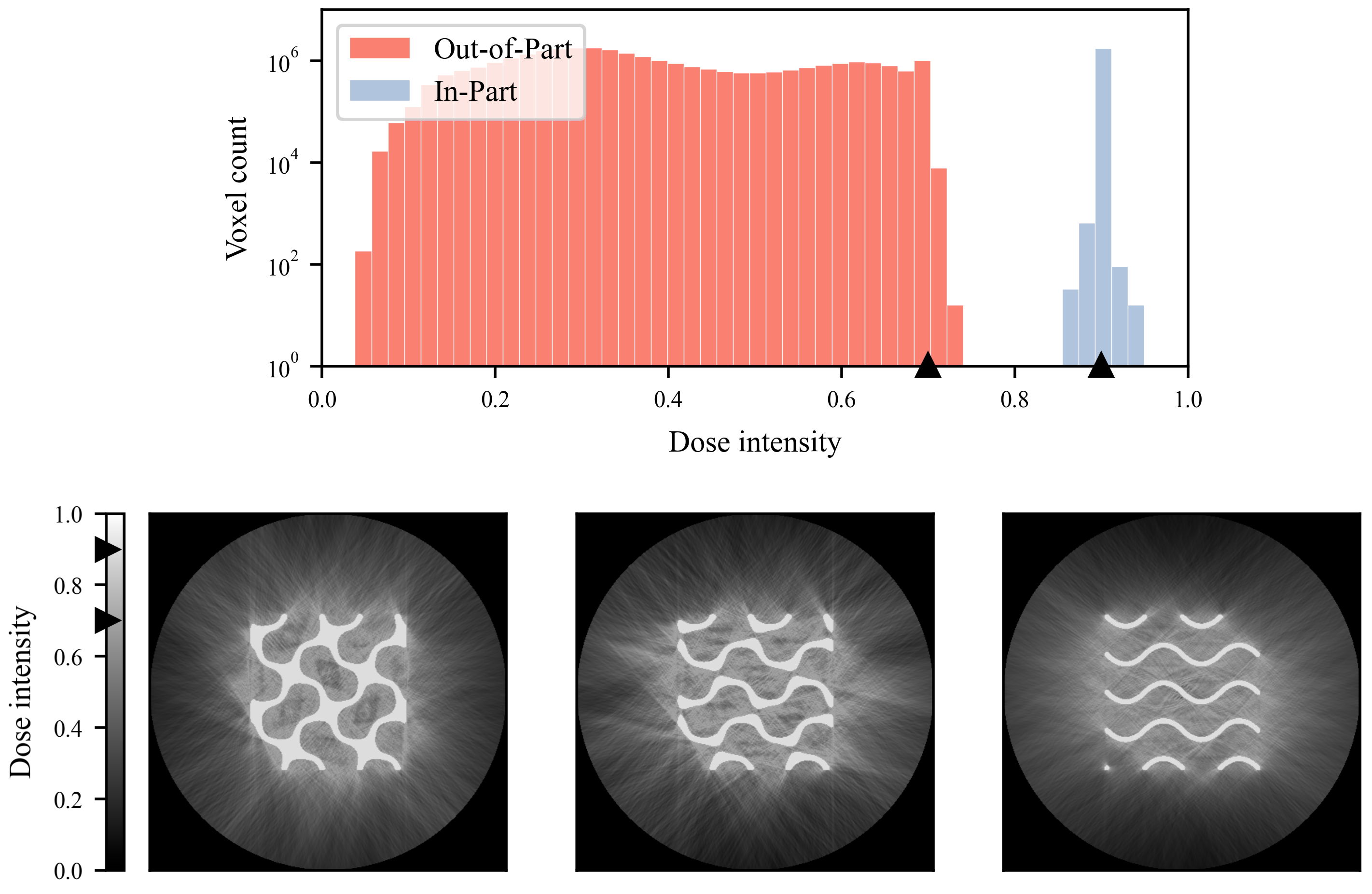}
    \caption{The histogram for all 198 slices of the 3D geometry (top) and the achieved dose profiles of slices at~$z = [14, 107, 197]$ (bottom, left-right) using~$\OSPW{=0.0}$ with default threshold pair~$\taulower = 0.70$ and~$\tauupper = 0.90$.}
    \label{fig:defaults3D-mixed}
\end{figure*}

\begin{figure*}
    \centering
    \includegraphics[]{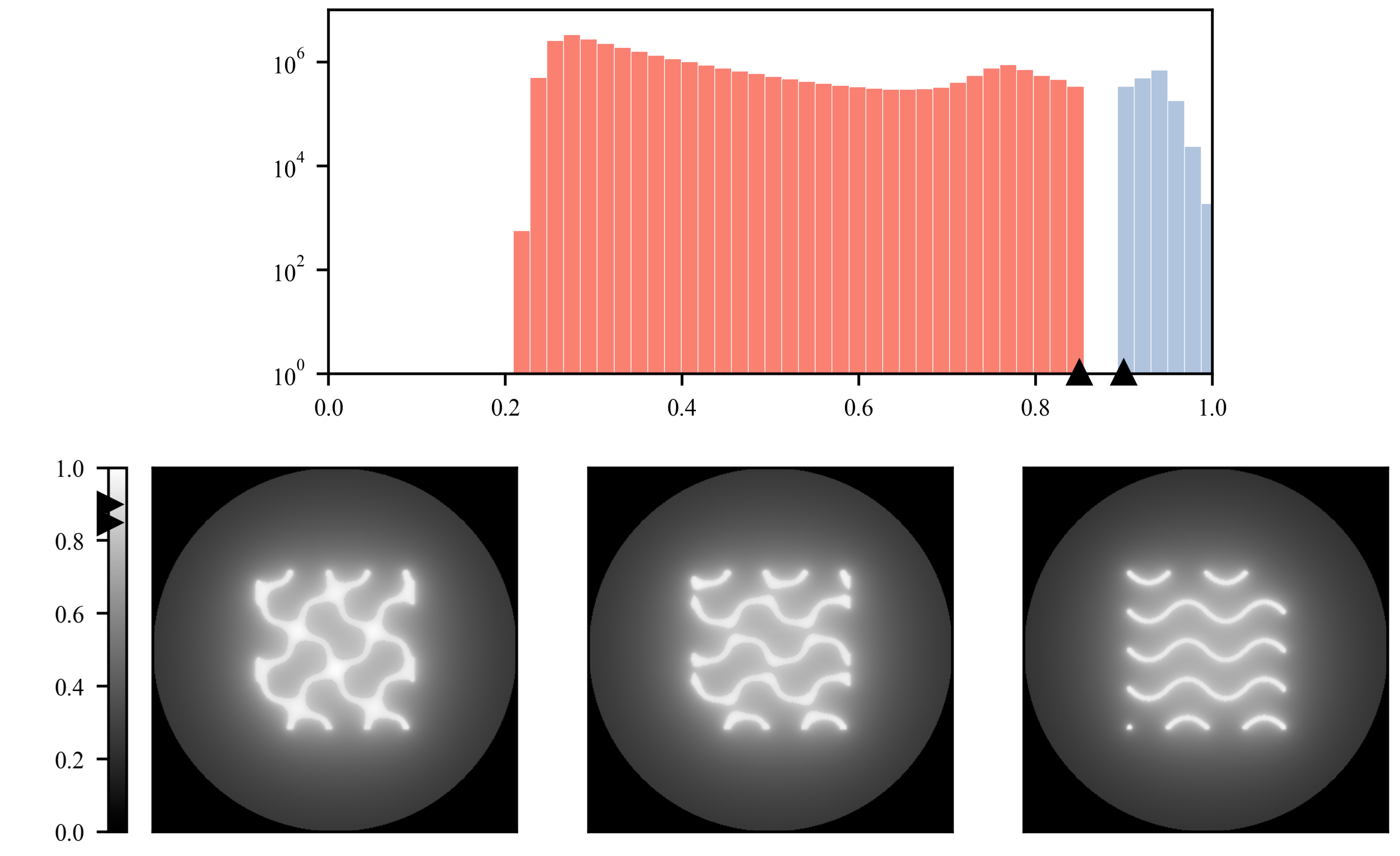}
    \caption{The histogram for all 198 slices of the 3D geometry (top) and the achieved dose profiles of slices at~$z = [14, 107, 197]$ (bottom, left-right) using OSMO with default threshold pair~$\taulower = 0.85$ and~$\tauupper = 0.90$.}
    \label{fig:defaults3D-OSMO}
\end{figure*}

\begin{figure}
    \centering
    \includegraphics[width=0.99\linewidth]{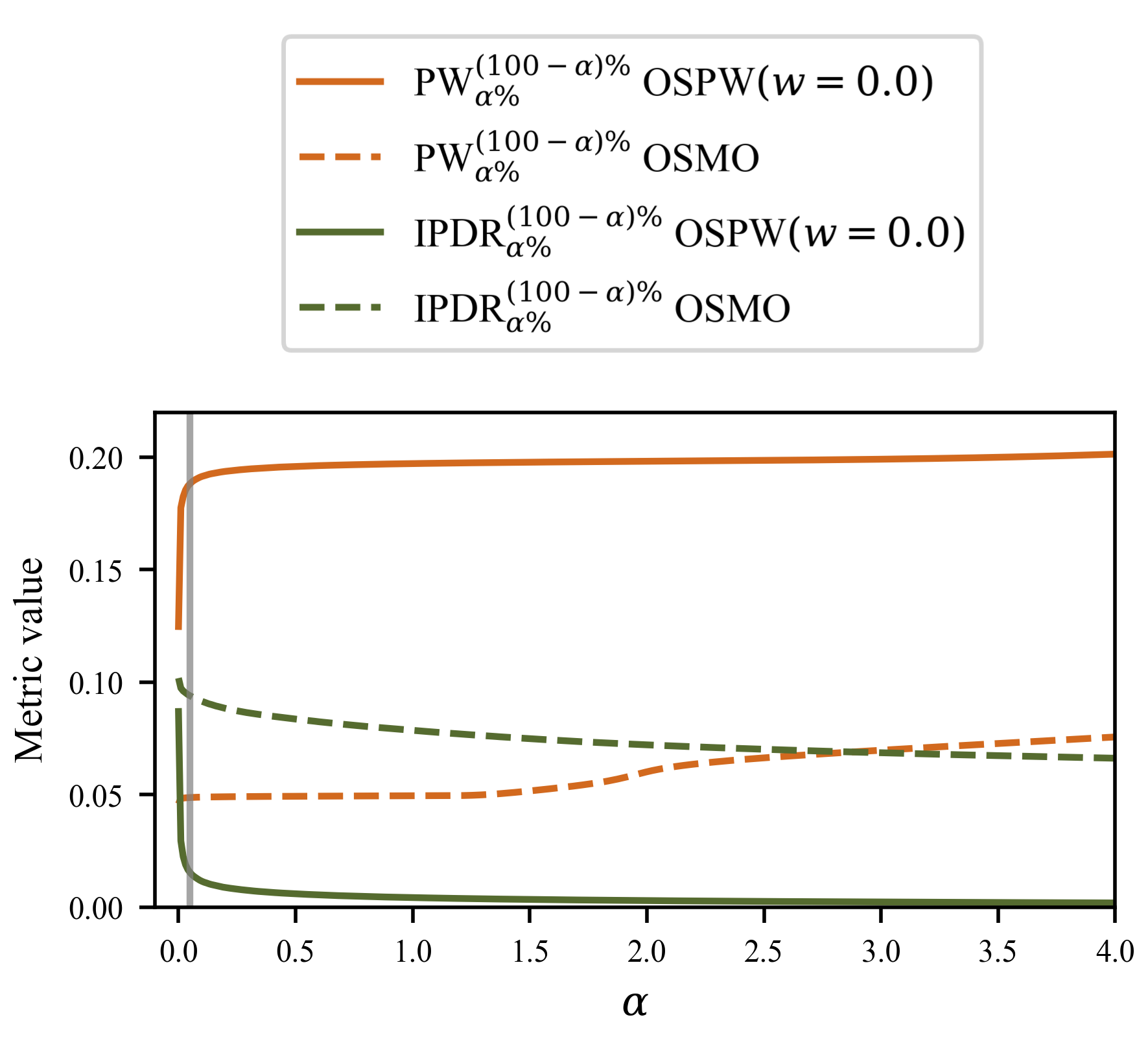}
    \caption{Metric values ($\PW$, $\IPDR$) for the full 3D Gyroid under increasing levels of percent data parametrised by~$\alpha$ removed from both extremes of both histograms. The vertical grey line indicates~$0.05\%$ of voxels each from in-part and out-of-part histograms not considered in the metric calculations. Recall that~$\PW$ should be maximised and~$\IPDR$ should be minimised. The default thresholds are used for~$\OSPW{=0.0}$ ($\taulower = 0.70$, $\tauupper = 0.90$) and OSMO ($\taulower = 0.85$, $\tauupper = 0.90$).}
    \label{fig:slice-by-slice}
\end{figure}

\begin{figure*}
    \begin{subfigure}{0.49\linewidth}
        \centering
        \includegraphics[]{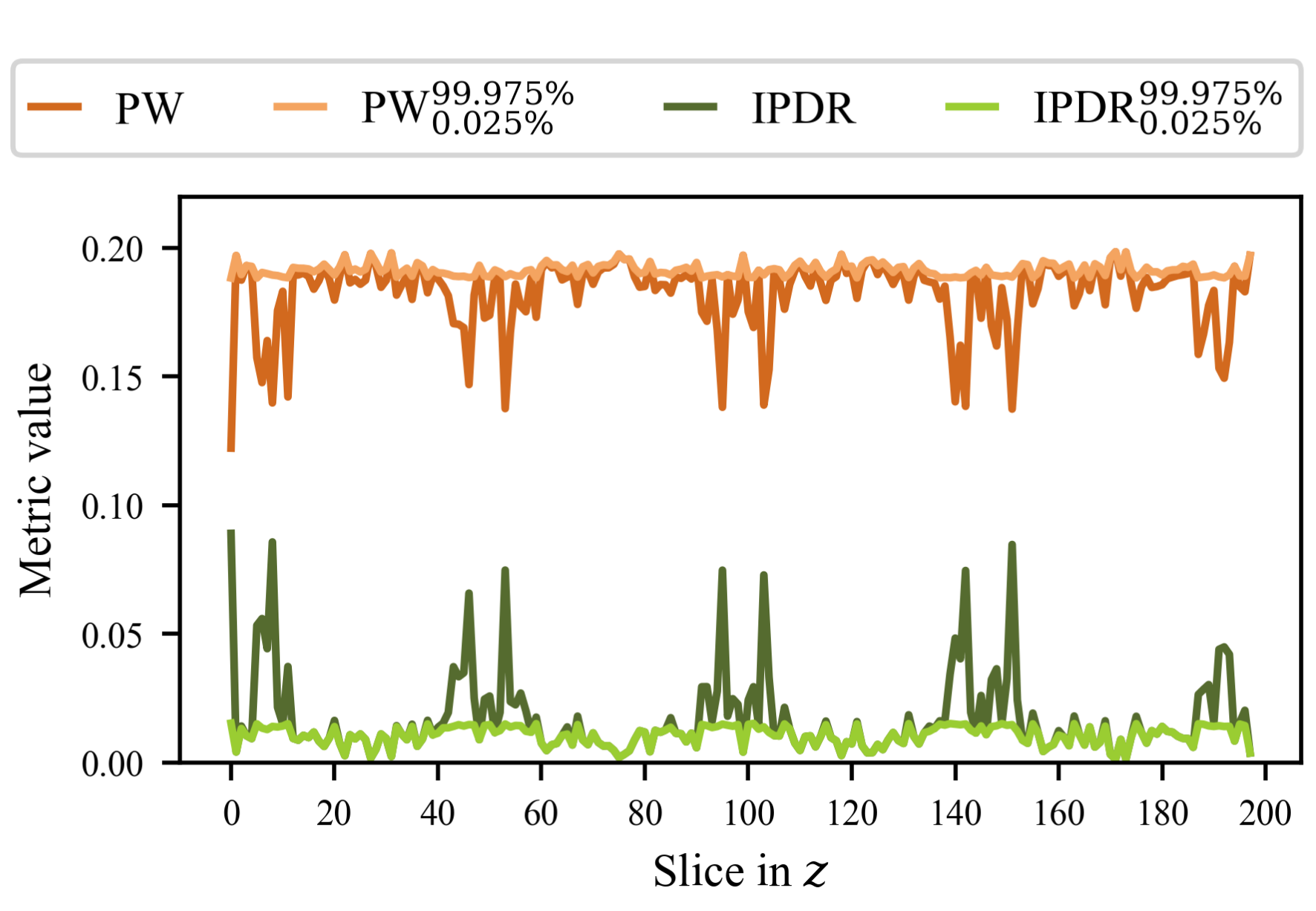}
        \caption{$\OSPW{=0.0}$ derived result.}
    \end{subfigure}
    \begin{subfigure}{0.49\linewidth}
        \centering
        \includegraphics[]{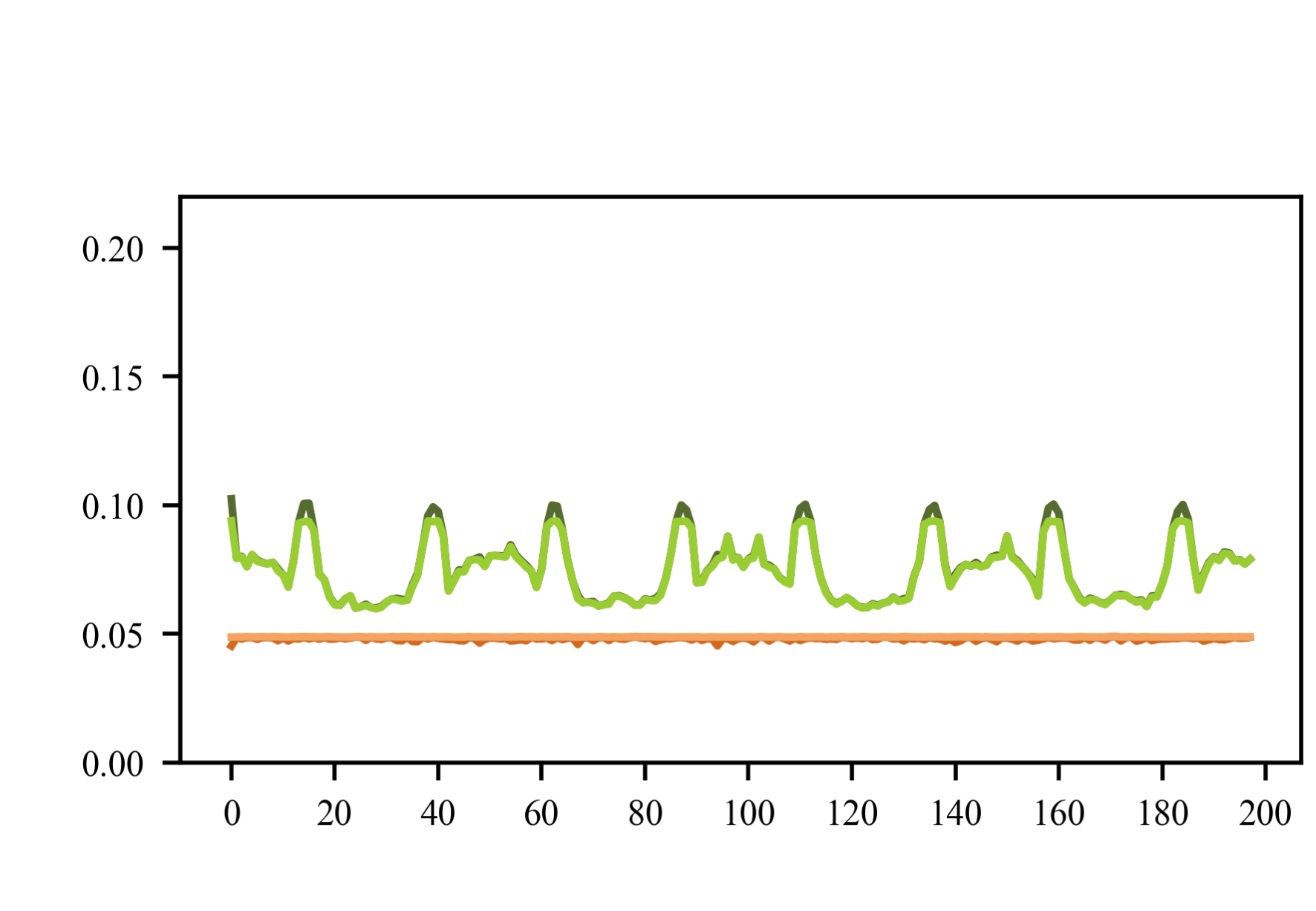}
        \caption{OSMO derived result.}
    \end{subfigure}
    \caption{Metric values taken from the full 3D light-energy dose profile, reported on a slice-by-slice basis. Recall, the subscript and superscripts relate to the lower and upper percentiles removed from the data, hence making the metric more stable to outliers; see Section~\ref{Section_method_metrics}.}
    \label{fig:discardMetrics}
\end{figure*}

\begin{figure*}
    \centering
    \includegraphics[]{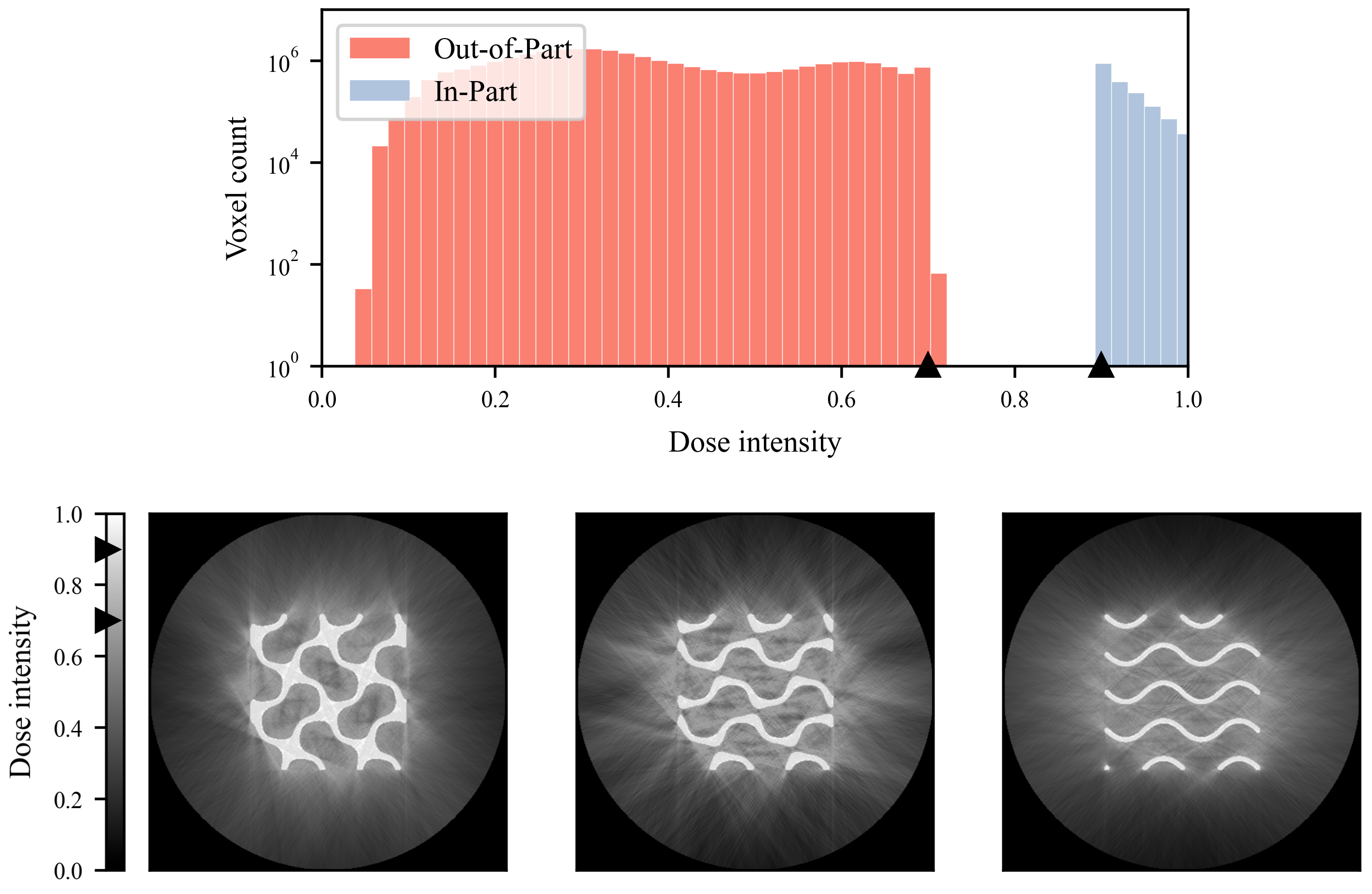}
    \caption{The histogram for all 198 slices of the 3D geometry (top) and the achieved dose profiles of slices at~$z = [14, 107, 197]$ (bottom, left-right) using~$\OSPW{=0.1}$ with default threshold pair~$\taulower = 0.70$ and~$\tauupper = 0.90$.}
    \label{fig:defaults3D-notSoOneSided}
\end{figure*}

As the projection geometry has been assumed to be parallel, extension to 3D is simple as each $x-y$ slice acts independently, but nonetheless important as printing is always a 3D exercise. The gyroid geometry (STL from~\cite{gyroid}) used to this effect here is shown in Figure~\ref{fig:geometry3D}; three of the 198~$x\text{-}y$ slices, which could each be considered as an individual geometry, are highlighted. These will be used to show achieved dose profiles from 1500 iterations. The results for~$\OSPW{=0.0}$ with default thresholds~$\taulower = 0.70$ and~$\tauupper = 0.90$ are shown in Figure~\ref{fig:defaults3D-mixed} and the result of OSMO with default thresholds~$\taulower = 0.85$ and~$\tauupper = 0.90$ are shown in Figure~\ref{fig:defaults3D-OSMO}. The~$\OSPW{=0.0}$ histogram shows a far larger~$\PW$ which could be expected due to the~$\OSPW{=0.0}$ default thresholds being set further apart than the OSMO default thresholds. Whilst, both results have a similar~$\IPDR$ ($0.09$ for~$\OSPW{=0.0}$ and~$0.11$ for OSMO), the mean value of achieved dose for voxels in~$I_\mathrm{in}$ is~$0.90$ for~$\OSPW{=0.0}$ and~$0.94$ for OSMO. This likely contributes to the brighter appearance of the in-part region of the OSMO achieved dose profile slice images compared to the~$\OSPW{=0.0}$ achieved dose profile slice images.

The in-part histogram of the~$\OSPW{=0.0}$ result could be described as having a bell-shaped distribution of dose intensity; a clear influence of the L2-norm penalty applied to these voxels. Revisiting the notion of metric robustness to outlier dose values, in Figure~\ref{fig:slice-by-slice}, increasing percentiles of voxels are discarded equally from top and bottom of both in-part and out-of-part histograms during~$\IPDR$ and~$\PW$ calculation.

Both of these metrics for the energy dose profile achieved by~$\OSPW{=0.0}$ experience a significant change when a very small percentile of voxels are discarded from the calculation; the vertical grey line marks~$0.05\%$ voxels each from in-part and out-of-part histograms not considered which is equally split to~$0.025\%$ at either end of each histogram. The otherwise relatively constant values can justify that removal of~$0.05\%$ voxels from in-part and out-of-part histograms for metric calculation is simply not including voxels with outlier dose values. This does not appear to be the case for the OSMO metrics.

Considering the metrics of the achieved dose (with no data removal) on a slice-by-slice basis,~$\OSPW{=0.0}$ shows far more variability in performance between slices~(Figure~\ref{fig:discardMetrics}a) compared to OSMO~(Figure~\ref{fig:discardMetrics}b). In Figure~\ref{fig:discardMetrics}a, it can be seen how closely linked~$\PW$ and~$\IPDR$ are, and that there are certain slices that perform significantly worse than the majority (note the periodic nature due to the~$(2\times2\times2)$ assembly of unit cells that compose the target geometry). In contrast, OSMO does well to maintain a near-constant~$\PW$ across all slices, and whilst there is inter-slice variation of~$\IPDR$, this is far lower than the inter-slice~$\IPDR$ variation from~$\OSPW{=0.0}$. Despite this, no slices of OSMO have a better~$\PW$ than any~$\OSPW{=0.0}$ slice, and almost all slices of~$\OSPW{=0.0}$ have a lower~$\IPDR$.

When~$0.05\%$ voxels each the in-part and out-of-part histograms are removed from the metric calculations, the inter-slice variation of~$\OSPW{=0.0}$ is highly dampened. Considering the full geometry (all 198 slices), there is a decrease in~$\IPDR$ from~$0.09$ to~$0.02$ and~$\PW$ increases from~$0.12$ to~$0.19$. Concurrently, OSMO shows minor changes to inter-slice variation of either metric, and for the full geometry neither~$\PW$ nor~$\IPDR$ improve from~$0.05$ and~$0.10$ respectively.

Finally, as discussed in Section~\ref{section_comparison_proposed_methods}, competing penalties may restrict exploration of possible outcomes during optimisation and relaxing the penalties could provide better results. This is assessed by using~$\OSPW{=0.1}$; the L2-norm penalty on voxels in~$I_\mathrm{in}$ is relaxed to a one-sided penalty with an additional penalty to limit high light-energy values. The result~(Figure~\ref{fig:defaults3D-notSoOneSided}) is better adherence to the thresholds compared to~$\OSPW{=0.0}$ which could be interpreted as the~$\OSPW{=0.0}$ penalty functions not being the best match for the gyroid geometry. The~$\IPDR$ increases slightly to~$0.11$ which is largely driven by~$\tauupper = 0.9$ but there is a big increase of~$\PW$ to~$0.18$, double the~$\OSPW{=0.0}$ result (when all voxels are used for metric calculation), showing that flexible penalty choice can result in significant improvement to the achieved dose profile.

\section{Conclusions}
\label{section_conclusions}

This work presents a critical analysis of penalty functions used in the creation of illumination plans for TVAM. The investigated penalty functions assumed a binary material model where voxels are classed as either belonging to the in-part region to be cured, or the out-of-part region and to remain liquid. Thus, curing occurs when the achieved dose received by a given voxel surpasses a threshold~$\tau$. The four penalty functions examined are:
\begin{enumerate}
	\item \textbf{L2N} uses the L2-norm fitted to the target image adjusted to encapsulate threshold information: the voxels in the out-of-part region should receive a low achieved dose of~$\taulower < \tau$ and the voxels in in-part region should receive a high achieved dose of~$\tauupper > \tau$,
	\item \textbf{OSP} uses more relaxed One-Sided Penalties which removes any penalty on voxels in the in-part region with an achieved dose greater than~$\tauupper$ and any penalty in the out-of-part region with an achieved dose less than~$\taulower$,
	\item \textbf{OSPW}$\bm{(w>0.0)}$ also uses one-sided penalties with an additional penalty to discourage the maximum achieved dose value at a width~$w$ above~$\tauupper$ with an aim to decrease~$\IPDR$ (first described by Wechsler\etal~\cite{waveOptical}),
	\item \textbf{OSPW}$\bm{(w>0.0)}$ uses a mixture of penalties with a one-sided penalty on voxels in the out-of-part region and an L2-norm penalty on voxels in the in-part region with the aim to preserve the large Process Window~($\PW$) of OSP and small In-Part Dose Range~($\IPDR$) of L2N.
\end{enumerate}

The dose profiles and resultant metrics achieved by these four approaches are first compared to one-another which suggests that~$\OSPW{=0.0}$ is the strongest performer as quantified by obtaining the lowest $\IPDR$ and highest $\PW$. Therefore,~$\OSPW{=0.0}$ is benchmarked against the well-known OSMO algorithm~\cite{OSMO} for three 2D target images where it is shown to consistently outperform OSMO in~$\PW$ and~$\IPDR$. This comparison selects optimal~$\taulower$ and~$\tauupper$ threshold pairs through a systematic parameter sweep which treats the thresholds as design parameters. The results show how the choice of threshold pairs greatly influences the metrics achieved by the dose profiles of the resultant illumination plans, and critically, that the formulation of the optimisation problem used in this work promotes a predictable metric response to changes in thresholds enabling an informed recommendation of default thresholds.

Finally, a 3D validation case is presented where~$\OSPW{=0.0}$ shows a marginal increase of~$\IPDR$ compared to OSMO and boasts a higher~$\PW$. It is also shown that for the selected geometry~$\PW$ performance can be boosted by using~$\OSPW{=0.1}$ penalties without sacrificing~$\IPDR$ thus making use of the flexibility of approach delivered by TVAM AID. By adopting the parameter sweep method for~$w$ (which was used to inform the choice of~$\taulower$ and~$\tauupper$), further insight could be gained into the effect of~$w$ on~$\IPDR$ and~$\PW$.

Whilst not yet validated experimentally, an advantage of this framework lies in its reproducibility and extensibility due to use of the Core Imaging Library~(CIL). Thus, performance of dose profiles currently achieved with one-sided penalties in~$\Omegaout$ could be further improved through inclusion of a regularisation term, already implemented in CIL, to the optimisation problem with the aim to improve achieved dose uniformity in~$\Omegaout$. The proposed approaches could be further customised by applying weights, a functionality also already supported by CIL, to change relative penalty severity between voxels in the in-part and out-of-part regions, or between specific voxels. In 3D, the flexibility of the framework could be leveraged on a slice-by-slice basis thus tailoring penalty functions to individual segments of the target image. More broadly, the scope of TVAM AID could be extended with a continuous material model permitting greyscale printing; a potential path to printing with functional materials.

\section*{CRediT Authorship Contribution Statement}
\textbf{NP}: Conceptualisation, Methodology, Formal analysis, Software, Writing - Original draft, Writing - Review \& editing, Visualisation. \textbf{RH}: Methodology, Software, Writing - Original draft, Writing - Review \& editing, Visualisation. \textbf{JS}:  Methodology, Writing - Review \& editing, Supervision, Project administration, Funding acquisition. \textbf{JSJ}: Conceptualisation, Methodology, Writing - Review \& editing, Supervision.

\section*{Acknowledgements}
The authors wish to acknowledge the support of the Collaborative Computation Project in Tomographic Imaging \href{https://gtr.ukri.org/projects?ref=EP%2FT026677%2F1}{(EPSRC grant EP/T026677/1)}. This work made use of computational support by CoSeC, the Computational Science Centre for Research Communities, through CCPi. \textbf{NP} and \textbf{JS} were supported by the Independent Research Fund Denmark (Contract No. 0171-00115B). \textbf{RH} and \textbf{JSJ} were supported by The Villum Foundation (Grant No. 25893). This research was funded in whole or in part by the Austrian Science Fund (FWF) 10.55776/\\F100800.

\appendix
\setcounter{figure}{0}
\makeatletter 
\renewcommand{\thefigure}{A\@arabic\c@figure}
\makeatother
\section{Additional Metric Maps} \label{app:heatMaps}
Below, in Figures~\ref{fig:dogbonesHeatMaps} to~\ref{fig:resolutionHeatMaps}, the metric colour maps for all three 2D geometries and the four investigated approaches (analogous to Figure~\ref{fig:parameterChoiceMixed}) show predictability in the metrics regardless of geometry or even choice of penalty functions. Further, the optimal threshold pair is consistently close to the default indicating the good choice of default thresholds.

\begin{figure*}
    \centering
    \begin{subfigure}{\linewidth}
        \centering
        \includegraphics{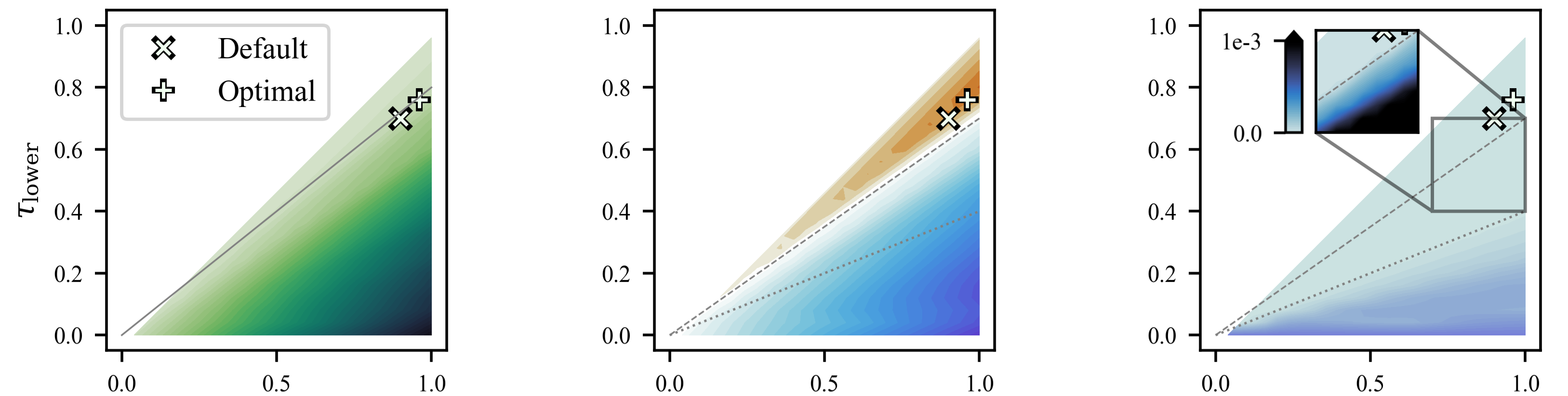}
        \caption{Colour maps for three critical metrics derived from the results of 325 threshold pairs using the L2N approach. The solid line is~$\taulower = 0.80\tauupper$ (left), the dashed line is~$\taulower = 0.70\tauupper$ (centre, right), and the dotted line is~$\taulower = 0.40\tauupper$ (centre, right).}
    \end{subfigure}
    \par \bigskip
    \begin{subfigure}{\linewidth}
        \centering
        \includegraphics{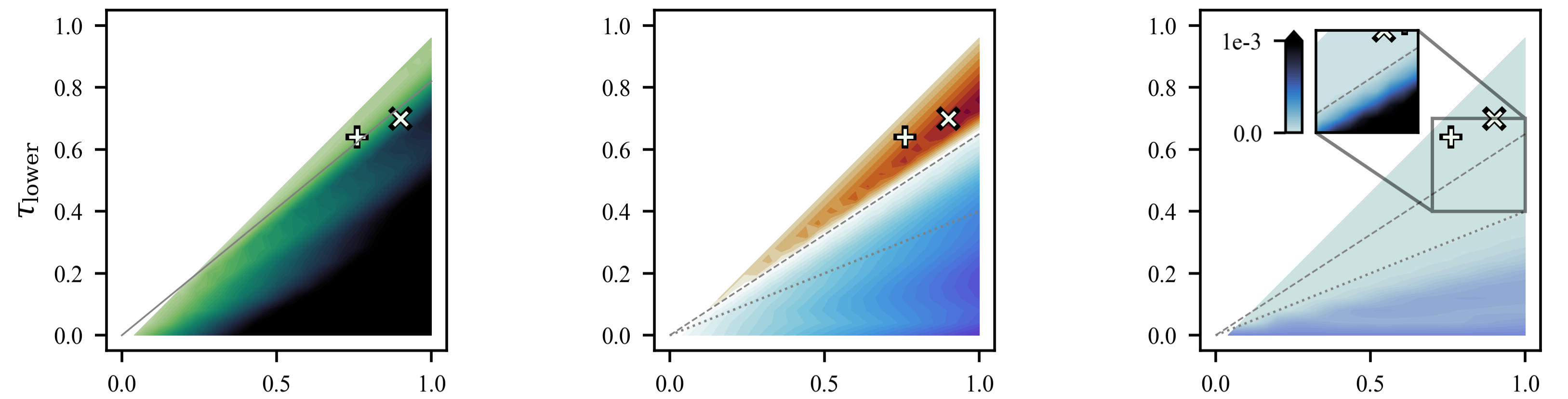}
        \caption{Colour maps for three critical metrics derived from the results of 325 threshold pairs using the OSP approach. The solid line is~$\taulower = 0.82\tauupper$ (left), the dashed line is~$\taulower = 0.65\tauupper$ (centre, right), and the dotted line is~$\taulower = 0.40\tauupper$ (centre, right).}
    \end{subfigure}
    \par \bigskip
    \begin{subfigure}{\linewidth}
        \centering
        \includegraphics{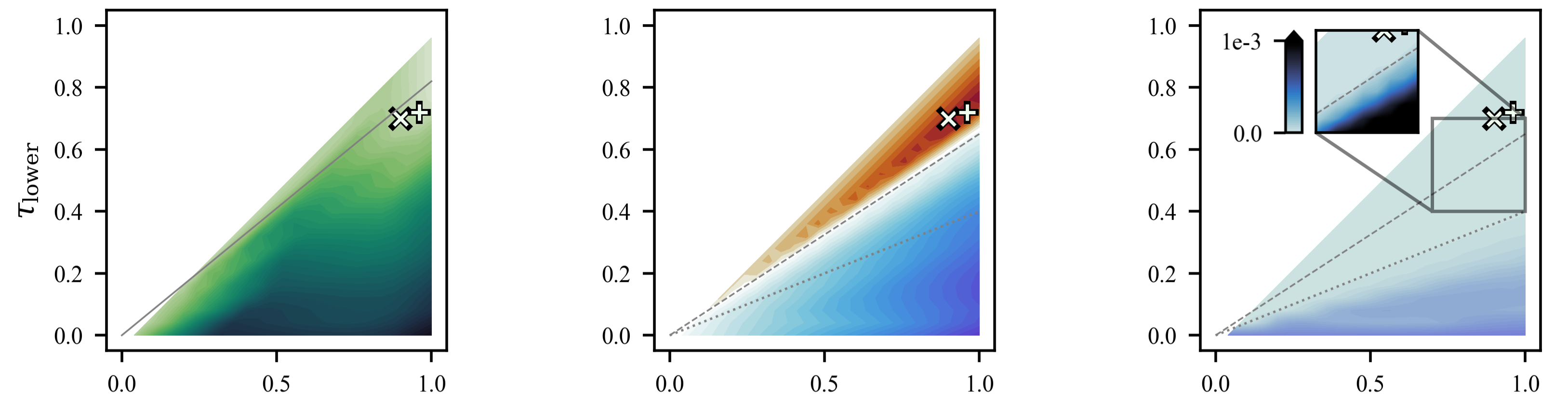}
        \caption{Colour maps for three critical metrics derived from the results of 325 threshold pairs using the~$\OSPW{> 0.0}$ approach where the width of no penalty in the penalty function varies with threshold such that~$w = 1 - \tauupper$. The solid line is~$\taulower = 0.82\tauupper$ (left), the dashed line is~$\taulower = 0.65\tauupper$ (centre, right), and the dotted line is~$\taulower = 0.40\tauupper$ (centre, right).}
    \end{subfigure}
    \par \bigskip
    \begin{subfigure}{\linewidth}
        \centering
        \includegraphics{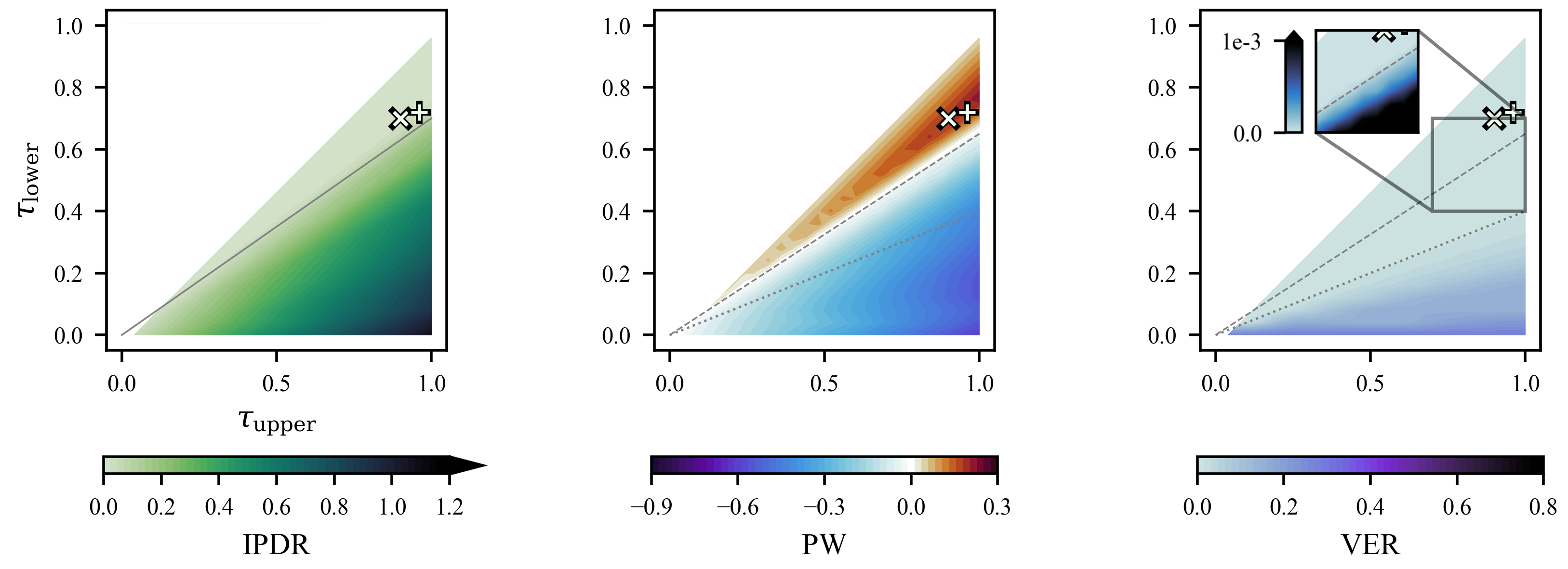}
        \caption{Colour maps for three critical metrics derived from the results of 325 threshold pairs using the~$\OSPW{=0.0}$ approach where the width of no penalty in the penalty function varies with threshold such that~$w = 1 - \tauupper$. The solid line is~$\taulower = 0.65\tauupper$ (left), the dashed line is~$\taulower = 0.75\tauupper$ (centre, right), and the dotted line is~$\taulower = 0.40\tauupper$ (centre, right).}
    \end{subfigure}
    \caption{Metric Colour maps for the DTU logo geometry for all investigated approaches.}
    \label{fig:dogbonesHeatMaps}
\end{figure*}

\begin{figure*}
    \centering
    \begin{subfigure}{\linewidth}
        \centering
        \includegraphics{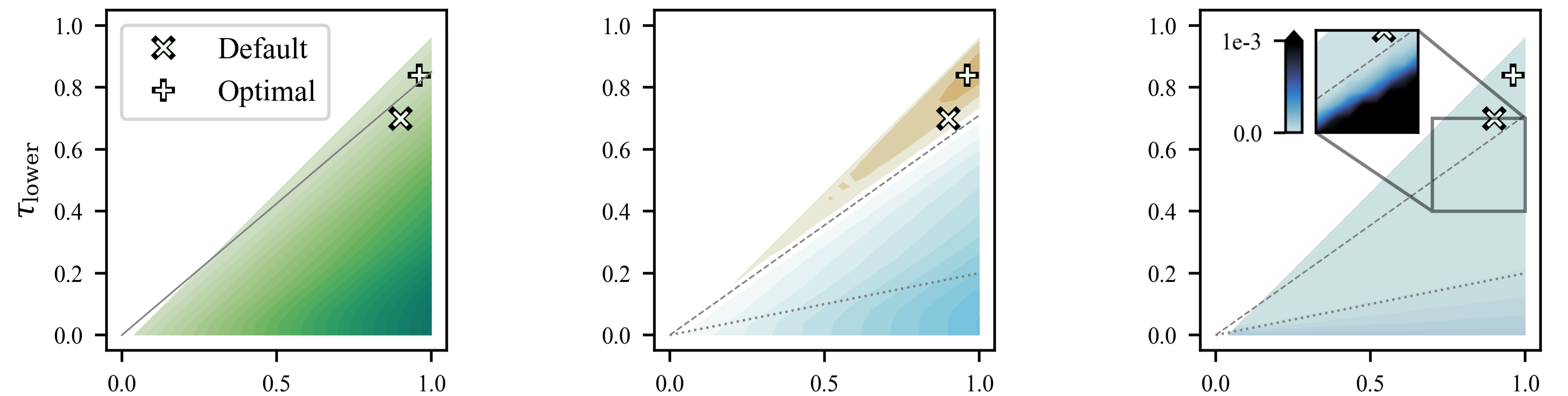}
        \caption{Colour maps for three critical metrics derived from the results of 325 threshold pairs using the L2N approach. The solid line is~$\taulower = 0.85\tauupper$ (left), the dashed line is~$\taulower = 0.71\tauupper$ (centre, right), and the dotted line is~$\taulower = 0.20\tauupper$ (centre, right).}
    \end{subfigure}
    \par \bigskip
    \begin{subfigure}{\linewidth}
        \centering
        \includegraphics{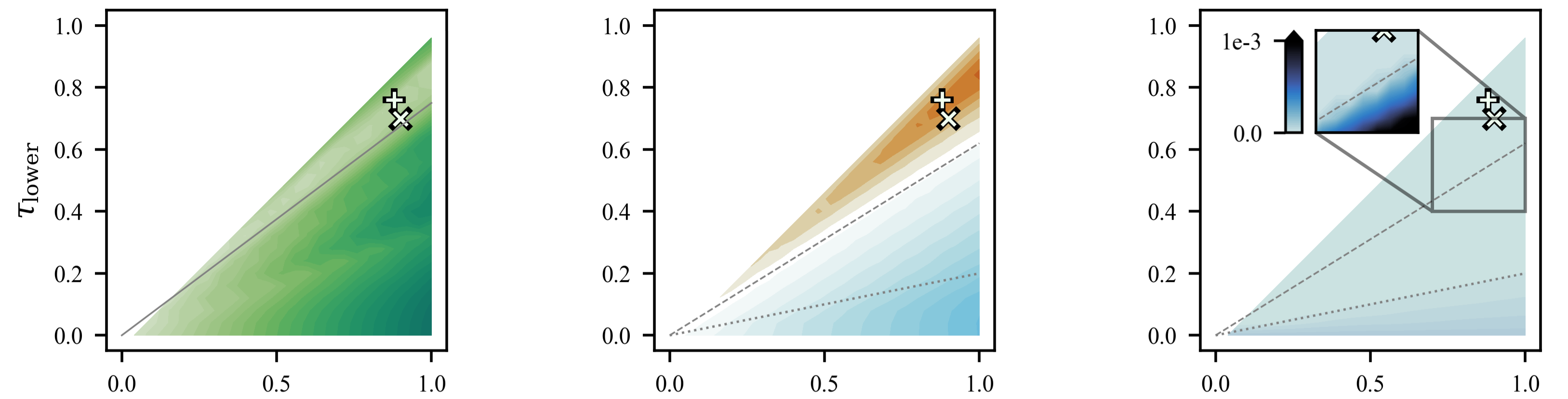}
        \caption{Colour maps for three critical metrics derived from the results of 325 threshold pairs using the OSP approach. The solid line is~$\taulower = 0.75\tauupper$ (left), the dashed line is~$\taulower = 0.62\tauupper$ (centre, right), and the dotted line is~$\taulower = 0.2\tauupper$ (centre, right).}
        
    \end{subfigure}
    \par \bigskip
    \begin{subfigure}{\linewidth}
        \centering
        \includegraphics{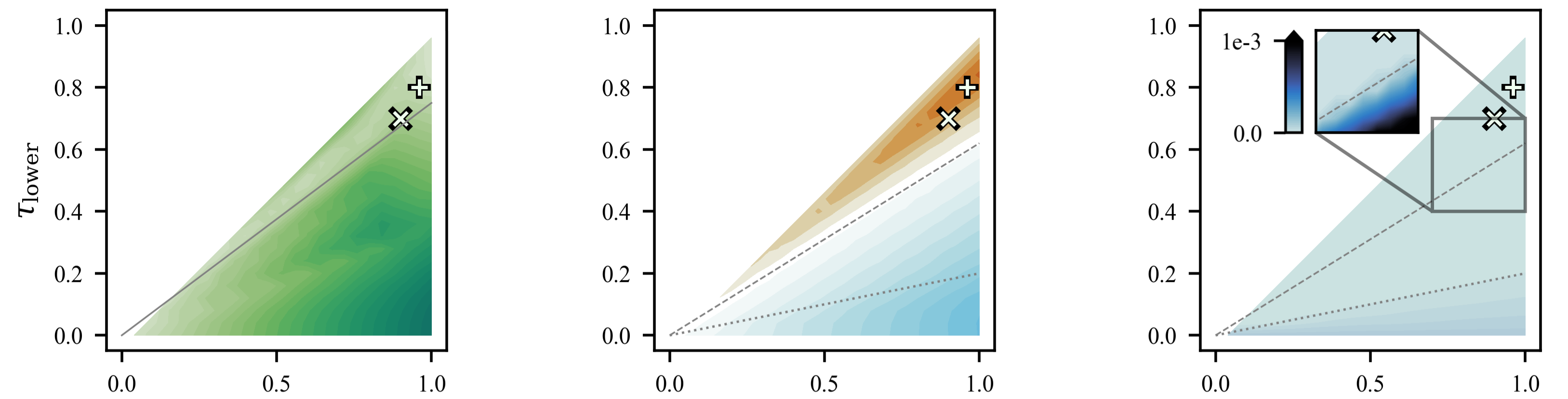}
        \caption{Colour maps for three critical metrics derived from the results of 325 threshold pairs using the~$\OSPW{>0.0}$ approach where the width of no penalty in the penalty function varies with threshold such that~$w = 1 - \tauupper$. The solid line is~$\taulower = 0.75\tauupper$ (left), the dashed line is~$\taulower = 0.62\tauupper$ (centre, right), and the dotted line is~$\taulower = 0.2\tauupper$ (centre, right).}
    \end{subfigure}
    \par \bigskip
    \begin{subfigure}{\linewidth}
        \centering
        \includegraphics{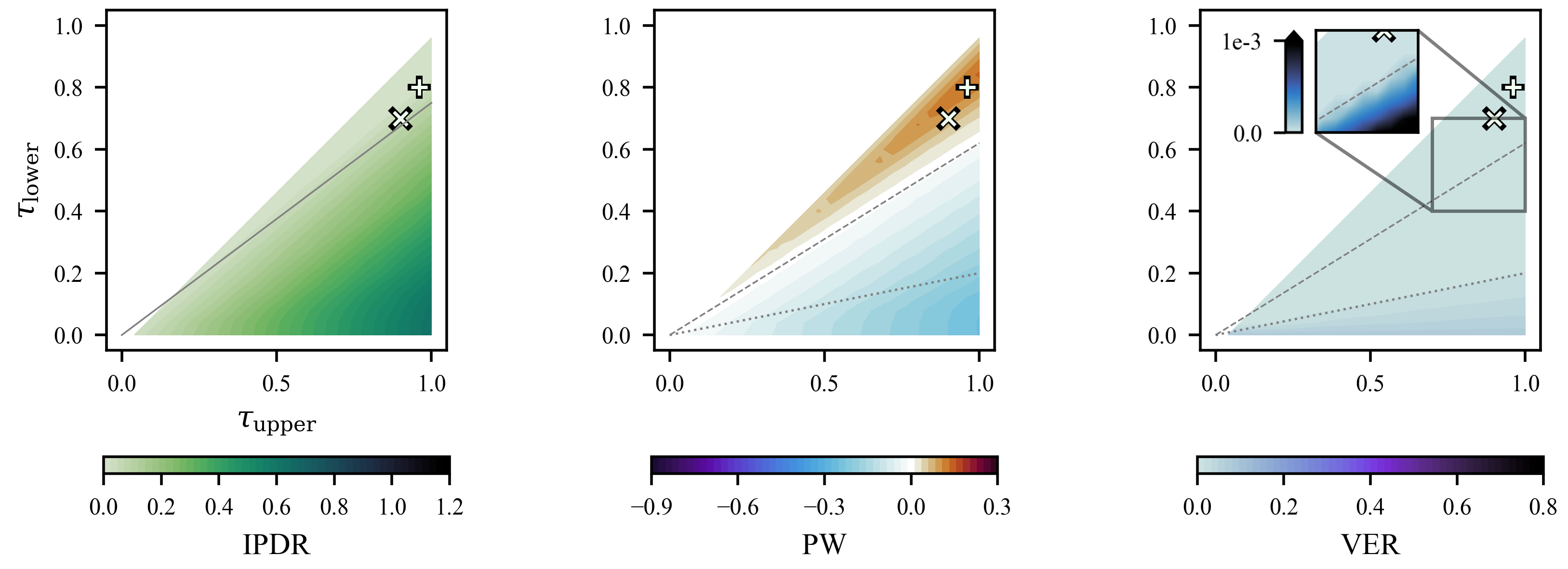}
        \caption{Colour maps for three critical metrics derived from the results of 325 threshold pairs using the~$\OSPW{=0.0}$ approach. The solid line is~$\taulower = 0.75\tauupper$ (left), the dashed line is~$\taulower = 0.62\tauupper$ (centre, right), and the dotted line is~$\taulower = 0.2\tauupper$ (centre, right).}
    \end{subfigure}
    \caption{Metric Colour maps for the disk geometry for all investigated approaches.}
    \label{fig:diskHeatMaps}
\end{figure*}

\begin{figure*}
    \centering
    \begin{subfigure}{\linewidth}
        \centering
        \includegraphics{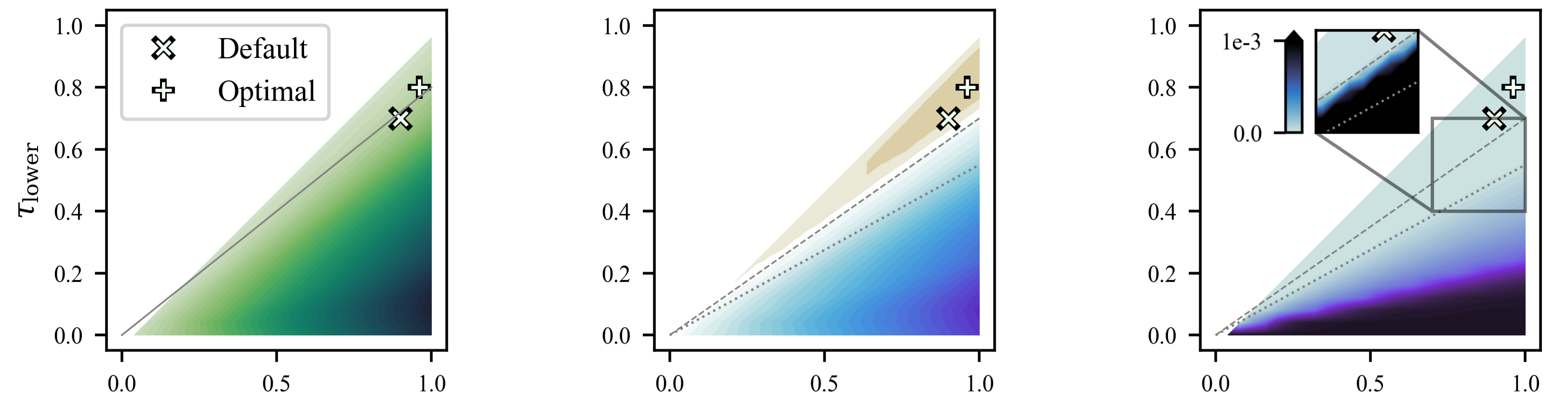}
        \caption{Colour maps for three critical metrics derived from the results of 325 threshold pairs using the L2N approach. The solid line is~$\taulower = 0.80\tauupper$ (left), the dashed line is~$\taulower = 0.70\tauupper$ (centre, right), and the dotted line is~$\taulower = 0.55\tauupper$ (centre, right).}
    \end{subfigure}
    \par \bigskip
    \begin{subfigure}{\linewidth}
        \centering
        \includegraphics{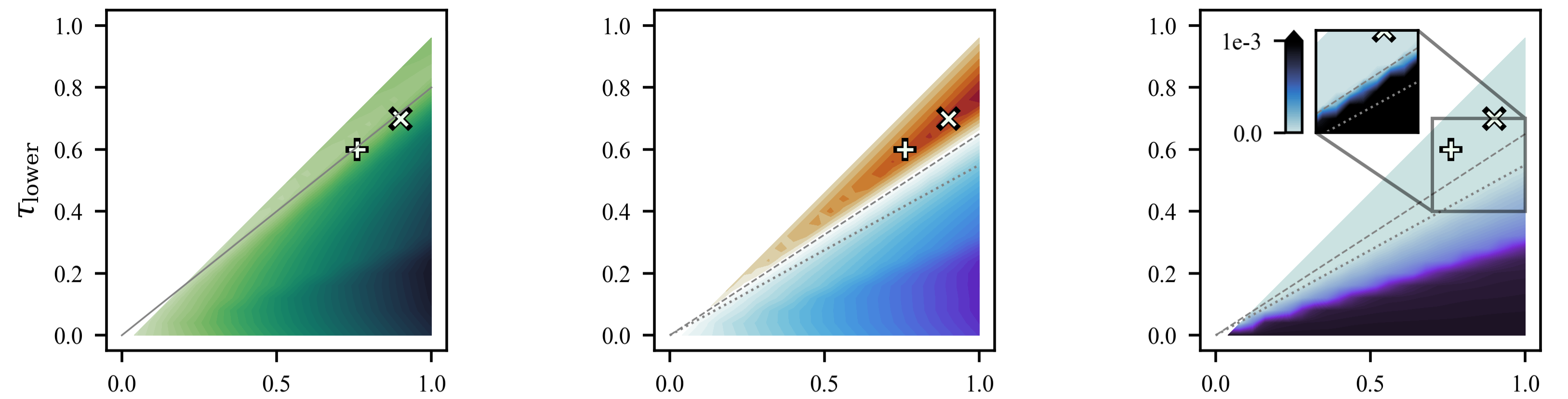}
        \caption{Colour maps for three critical metrics derived from the results of 325 threshold pairs using the OSP approach. The solid line is~$\taulower = 0.8\tauupper$ (left), the dashed line is~$\taulower = 0.65\tauupper$ (centre, right), and the dotted line is~$\taulower = 0.55\tauupper$ (centre, right).}
    \end{subfigure}
    \par \bigskip
    \begin{subfigure}{\linewidth}
        \centering
        \includegraphics{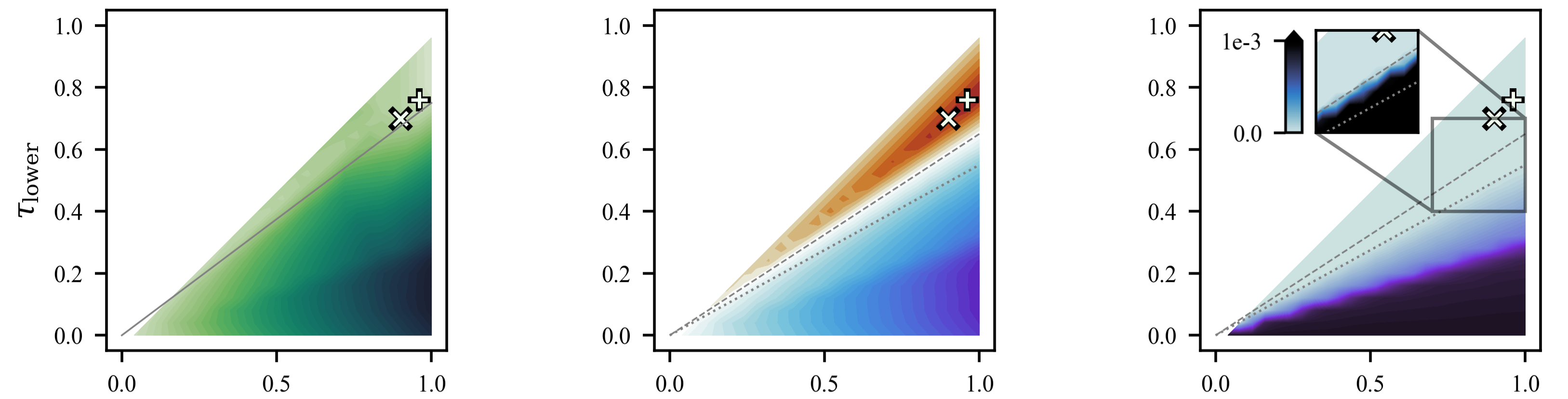}
        \caption{Colour maps for three critical metrics derived from the results of 325 threshold pairs using the~$\OSPW{>0.0}$ approach where the width of no penalty in the penalty function varies with threshold such that~$w = 1 - \tauupper$. The solid line is~$\taulower = 0.75\tauupper$ (left), the dashed line is~$\taulower = 0.65\tauupper$ (centre, right), and the dotted line is~$\taulower = 0.55\tauupper$ (centre, right).}
    \end{subfigure}
    \par \bigskip
    \begin{subfigure}{\linewidth}
        \centering
        \includegraphics{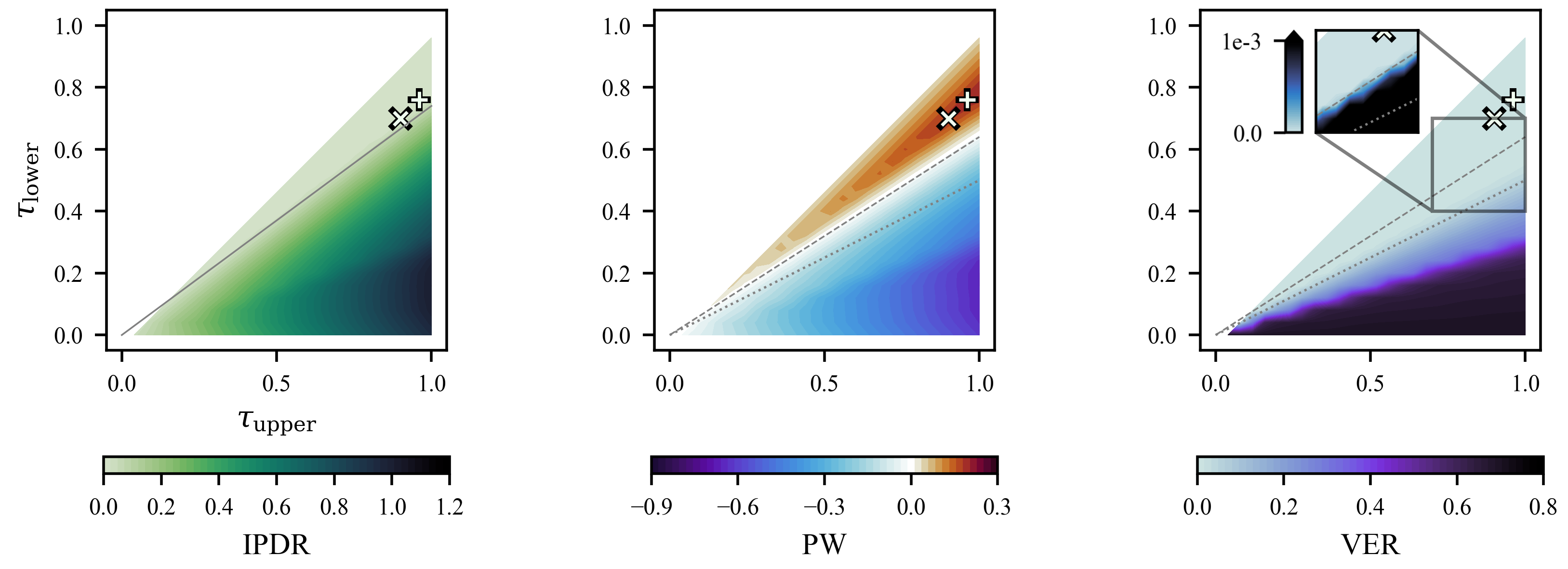}
        \caption{Colour maps for three critical metrics derived from the results of 325 threshold pairs using the~$\OSPW{=0.0}$ approach. The solid line is~$\taulower = 0.\tauupper$ (left), the dashed line is~$\taulower = \tauupper$ (centre, right), and the dotted line is~$\taulower = \tauupper$ (centre, right).}
    \end{subfigure}
    \caption{Metric Colour maps for the resolution test geometry for all investigated approaches.}
    \label{fig:resolutionHeatMaps}
\end{figure*}

\bibliographystyle{elsarticle-num} 
\bibliography{refs}

@article{CAL,
    title   = {Volumetric Additive Manufacturing via Tomographic Reconstruction},
    author  = {Kelly, Brett E. and
               Bhattacharya, Indrasen and
               Heidari, Hossein and
               Shusteff, Maxim and
               Spadaccini, Christopher M. and
               Taylor, Hayden K.},
    journal = {Science},
    volume  = {363},
    number  = {6431},
    pages   = {1075--1079},
    year    = {2019},
    doi     = {10.1126/science.aau7114}
}

@article{VirtualVam,
    title   = {Virtual Volumetric Additive Manufacturing {(VirtualVAM)}},
    author  = {Weisgraber, Todd H. and
               de Beer, Martin P. and
               Huang, Sijia and
               Karnes, John J. and
               Cook, Caitlyn C. and
               Shusteff, Maxim},
    journal = {Advanced Materials Technologies},
    volume  = {8},
    number  = {23},
    pages   = {2301054},
    year    = {2023},
    doi     = {10.1002/admt.202301054}
}

@article{shusteff2017one,
  title     = {One-step Volumetric Additive Manufacturing of Complex Polymer Structures},
  author    = {Shusteff, Maxim and
               Browar, Allison EM and
               Kelly, Brett E and
               Henriksson, Johannes and
               Weisgraber, Todd H and
               Panas, Robert M and
               Fang, Nicholas X and
               Spadaccini, Christopher M},
  journal   = {Science Advances},
  volume    = {3},
  number    = {12},
  pages     = {eaao5496},
  year      = {2017},
  doi       = {10.1126/sciadv.aao5496}
}

@article{loterie,
    title   = {High-Resolution Tomographic Volumetric Additive Manufacturing},
    author  = {Loterie, Damien and
               Delrot, Paul and
               Moser, Christophe},
    journal = {Nature Communications},
    volume  = {11},
    number  = {1},
    pages   = {852},
    year    = {2020},
    doi     = {10.1038/s41467-020-14630-4}
}

@article{bernal2019volumetric,
  title     = {Volumetric bioprinting of complex living-tissue constructs within seconds},
  author    = {Bernal, Paulina Nu{\~n}ez and
               Delrot, Paul and
               Loterie, Damien and
               Li, Yang and
               Malda, Jos and
               Moser, Christophe and
               Levato, Riccardo},
  journal   = {Advanced Materials},
  volume    = {31},
  number    = {42},
  pages     = {1904209},
  year      = {2019},
  doi       = {10.1002/adma.201904209}
}

@article{monzon2017anisotropy,
  title     = {Anisotropy of photopolymer parts made by digital light processing},
  author    = {Monz{\'o}n, Mario and
               Ortega, Zaida and
               Hern{\'a}ndez, Alba and
               Paz, Rub{\'e}n and
               Ortega, Fernando},
  journal   = {Materials},
  volume    = {10},
  number    = {1},
  pages     = {64},
  year      = {2017},
  doi       = {10.3390/ma10010064}
}

@article{bikas2016additive,
  title     = {Additive manufacturing methods and modelling approaches: a critical review},
  author    = {Bikas, Harry and
               Stavropoulos, Panagiotis and
               Chryssolouris, George},
  journal   = {The International Journal of Advanced Manufacturing Technology},
  volume    = {83},
  number    = {1},
  pages     = {389--405},
  year      = {2016},
  doi       = {10.1007/s00170-015-7576-2}
}

@article{moroni2018biofabrication,
  title     = {Biofabrication: a guide to technology and terminology},
  author    = {Moroni, Lorenzo and
               Boland, Thomas and
               Burdick, Jason A and
               De Maria, Carmelo and
               Derby, Brian and
               Forgacs, Gabor and
               Groll, J{\"u}rgen and
               Li, Qing and
               Malda, Jos and
               Mironov, Vladimir A and
               others},
  journal   = {Trends in Biotechnology},
  volume    = {36},
  number    = {4},
  pages     = {384--402},
  year      = {2018},
  doi       = {10.1016/j.tibtech.2017.10.015}
}

@book{natterer2001mathematics,
  title     = {The mathematics of computerized tomography},
  author    = {Natterer, Frank},
  year      = {2001},
  publisher = {Society for Industrial and Applied Mathematics}
}

@book{deans2007radon,
  title     = {The Radon transform and some of its applications},
  author    = {Deans, Stanley R},
  year      = {2007},
  publisher = {Courier Corporation}
}

@article{li2024tomographic,
  title     = {Tomographic projection optimization for volumetric additive manufacturing with general band constraint {L}p-norm minimization},
  author    = {Li, Chi Chung and
               Toombs, Joseph T and
               Taylor, Hayden K and
               Wallin, Thomas J},
  journal   = {Additive Manufacturing},
  volume    = {94},
  pages     = {104447},
  year      = {2024},
  doi       = {10.1016/j.addma.2024.104447}
}

@article{highFidelity,
    title   = {High Fidelity Volumetric Additive Manufacturing},
    author  = {Bhattacharya, Indrasen and
               Toombs, Joseph T. and
               Taylor, Hayden K.},
    journal = {Additive Manufacturing},
    volume  = {47},
    number  = {},
    pages   = {102299},
    year    = {2021},
    doi     = {10.1016/j.addma.2021.102299}
}

@article{OSMO,
    title   = {Object-space Optimization of Tomographic Reconstructions for Additive Manufacturing},
    author  = {Charles M. Rackson and
               Kyle M. Champley and
               Joseph T. Toombs and
               Erika J. Fong and
               Vishal Bansal and
               Hayden K. Taylor and
               Maxim Shusteff and
               Robert R. McLeod},
    journal = {Additive Manufacturing},
    volume  = {48},
    pages   = {102367},
    year    = {2021},
    doi     = {10.1016/j.addma.2021.102367}
}

@article{rayDistortion,
    title   = {Correcting Ray Distortion in Tomographic Additive Manufacturing},
    author  = {Antony Orth and
               Kathleen L. Sampson and
               Kayley Ting and
               Jonathan Boisvert and
               Chantal Paquet},
    journal = {Optics Express},
    volume  = {29},
    number  = {7},
    pages   = {11037--11054},
    year    = {2021},
    doi     = {10.1364/OE.419795}
}

@article{murphy20143d,
  title     = {{3D} bioprinting of tissues and organs},
  author    = {Murphy, Sean V and
               Atala, Anthony},
  journal   = {Nature Biotechnology},
  volume    = {32},
  number    = {8},
  pages     = {773--785},
  year      = {2014},
  doi       = {10.1038/nbt.2958 }
}

@article{sandstrom2016non,
  title     = {The non-disruptive emergence of an ecosystem for {3D} Printing -- {I}nsights from the hearing aid industry's transition 1989--2008},
  author    = {Sandstr{\"o}m, Christian G},
  journal   = {Technological Forecasting and Social Change},
  volume    = {102},
  pages     = {160--168},
  year      = {2016},
  doi       = {10.1016/j.techfore.2015.09.006}
}

@article{martin20173d,
  title     = {3D printing of high-strength aluminium alloys},
  author    = {Martin, John H and
               Yahata, Brennan D and
               Hundley, Jacob M and
               Mayer, Justin A and
               Schaedler, Tobias A and
               Pollock, Tresa M},
  journal   = {Nature},
  volume    = {549},
  number    = {7672},
  pages     = {365--369},
  year      = {2017},
  doi       = {10.1038/nature23894}
}

@article{giannatsis2009additive,
  title     = {Additive Fabrication Technologies Applied to Medicine and Health Care: A Review},
  author    = {Giannatsis, J and Dedoussis, V},
  journal   = {The International Journal of Advanced Manufacturing Technology},
  volume    = {40},
  number    = {1},
  pages     = {116--127},
  year      = {2009},
  doi       = {Springer}
}

@article{frazier2014metal,
  title     = {Metal Additive Manufacturing: A Review},
  author    = {Frazier, William E},
  journal   = {Journal of Materials Engineering and Performance},
  volume    = {23},
  number    = {6},
  pages     = {1917--1928},
  year      = {2014},
  doi       = {10.1007/s11665-014-0958-z}
}

@article{newGermanVAMGroup,
    title   = {An Easy-to-build, Accessible Volumetric 3D Printer Based on a Liquid Crystal Display for Rapid Resin Development},
    author  = {Silvio Tisato and
               Grace Vera and
               Akshaya Mani and
               Timothy Chase and
               Dorothea Helmer},
    journal = {Additive Manufacturing},
    volume  = {87},
    number  = {},
    pages   = {104232},
    year    = {2024},
    doi     = {10.1016/j.addma.2024.104232}
}

@article{newGermanVAMGroup2,
    title   = {Additive manufacturing of multi-material and hollow structures by Embedded Extrusion-Volumetric Printing},
    author  = {Silvio Tisato and
               Grace Vera and
               Qingchuan Song and
               Niloofar Nekoonam and
               Dorothea Helmer},
    journal = {Nature Communications},
    volume  = {16},
    number  = {},
    pages   = {6730},
    year    = {2025},
    doi     = {10.1038/s41467-025-62057-6}
}

@article{edgeEnhanced,
    title   = {Edge-enhanced object-space model optimization of tomographic reconstructions for additive manufacturing},
    author  = {Zhang, Yanchao and
               Liu, Minzhe and
               Liu, Hua and
               Gao, Ce and
               Jia, Zhongqing and
               Zhai, Ruizhan},
    journal = {Micromachines},
    volume  = {14},
    number  = {7},
    pages   = {1362},
    year    = {2023},
    doi     = {10.3390/mi14071362}
}

@article{CILpaper,
    title   = {{Core Imaging Library - Part I}: A Versatile {Python} Framework for Tomographic Imaging},
    author  = {J{\o}rgensen, Jakob S. and
               Ametova, Evelina and
               Burca, Genoveva and
               Fardell, Gemma and
               Papoutsellis, Evangelos and
               Pasca, Edoardo and
               Thielemans, Kris and
               Turner, Martin and
               Warr, Ryan and
               Lionheart, William R. B. and
               Withers, Philip J.},
    journal = {Philosophical Transactions of the Royal Society {A}},
    volume  = {379},
    number  = {2204},
    pages   = {20200192},
    year    = {2021},
    doi     = {10.1098/rsta.2020.0192}
}

@misc{CILcode,
    title     = {{Core Imaging Library (CIL)}},
    author    = {Pasca, Edoardo and
                 J{\o}rgensen, Jakob S. and
                 Papoutsellis, Evangelos and
                 Ametova, Evelina and
                 Fardell, Gemma and
                 Thielemans, Kris and
                 Murgatroyd, Laura and
                 Duff, Margaret and
                 da Costa-Luis, Casper and
                 Roberts, Hannah and
                 Sugic, Danica},
    note      = {Version 24.2.0},
    year      = {2024},
    publisher = {Zenodo},
    doi       = {10.5281/zenodo.13918923}
}

@article{FISTA,
    title   = {A Fast Iterative Shrinkage-Thresholding Algorithm for Linear Inverse Problems},
    author  = {Beck, Amir and
              Teboulle, Marc},
    journal = {SIAM Journal on Imaging Sciences},
    volume  = {2},
    number  = {1},
    pages   = {183-202},
    year    = {2009},
    doi     = {10.1137/080716542}
}

@article{highlyTunable,
    title   = {Highly Tunable Thiol-Ene Photoresins for Volumetric Additive Manufacturing},
    author  = {Cook, Caitlyn C. and
               Fong, Erika J. and
               Schwartz, Johanna J. and
               Porcincula, Dominique H. and
               Kaczmarek, Allison C. and
               Oakdale, James S. and
               Moran, Bryan D. and
               Champley, Kyle M. and
               Rackson, Charles M. and
               Muralidharan, Archish and
               others},
    journal = {Advanced Materials},
    volume  = {32},
    number  = {47},
    pages   = {2003376},
    year    = {2020},
    doi     = {10.1002/adma.202003376}
}

@article{inverseRendering,
    title   = {Inverse Rendering for Tomographic Volumetric Additive Manufacturing},
    author  = {Baptiste, Nicolet and
               Wechsler, Felix and
               Madrid-Wolff, Jorge and
               Moser, Christophe and
               Wenzel, Jakob},
    journal = {{ACM} Transactions on Graphics},
    volume  = {43},
    number  = {6},
    pages   = {12{-ART}228},
    year    = {2024},
    doi     = {10.1145/3687924}
}

@article{3projectors,
    title   = {Printing of low-viscosity materials using tomographic additive manufacturing},
    author  = {Webber, Daniel and
               Orth, Antony and
               Vidyapin, Victor and
               Zhang, Yujie and
               Picard, Michel and
               Liu, David and
               Sampson, Kathleen L and
               Lacelle, Thomas and
               Paquet, Chantal and
               Boisvert, Jonathan},
    journal = {Additive Manufacturing},
    volume  = {94},
    number  = {},
    pages   = {104480},
    year    = {2024},
    doi     = {10.1016/j.addma.2024.104480}
}

@misc{gyroid,
    author = {Seth Moczydlowski},
    title  = {Gyroid Cube},
    year   = {2015},
    url    = {https://www.thingiverse.com/thing:757884},
    note   = {Accessed: 27/11/2025}
}

@misc{projectGithub,
    author = {Nicole Pellizzon},
    title  = {TVAM-AID},
    year   = {2025},
    note   = {},
    url    = {https://github.com/DTU-VAM/TVAM-AID}
}

@misc{OSMOCode,
    author  = {Joseph Toombs and
              Chi Chung Li},
    license = {GPL-3.0},
    title   = {{VAMToolbox}},
    year    = {2023},
    note    = {Version 2.0.0},
    url     = {https://github.com/computed-axial-lithography/VAMToolbox}
}

@misc{DTULogo,
    author = {{Teknologihistorie DTU}},
    title  = {{DTU}'s Navn og Logo},
    note   = {Accessed: 02/12/2025},
    url    = {https://historie.dtu.dk/dtus-historie/navn-og-logo}
}

@article{waveOptical,
    title   = {Wave Optical Model for Tomographic Volumetric Additive Manufacturing},
    author  = {Felix Wechsler and
               Carlo Gigli and
               Jorge Madrid-Wolff and
               Christophe Moser},
    journal = {Optics Express},
    volume  = {32},
    number  = {8},
    pages   = {14705},
    year    = {2024},
    doi     = {10.1364/OE.521322}
}

@article{webber2023versatile,
    title        = {Versatile volumetric additive manufacturing with 3D ray tracing},
    author       = {Webber, Daniel and Zhang, Yujie and Picard, Michel and Boisvert, Jonathan and Paquet, Chantal and Orth, Antony},
    journal      = {Optics Express},
    volume       = {31},
    number       = {4},
    pages        = {5531--5546},
    year         = {2023},
    publisher    = {Optica Publishing Group},
    doi          = {10.1364/OE.481318}
}

\end{document}